\documentclass[11pt, reqno]{article}
\usepackage{amsfonts}
\usepackage{amsmath}
\usepackage{amsthm}
\usepackage{amssymb}
\usepackage{lastpage}
\usepackage{fancyhdr}
\usepackage{fancybox}
\usepackage{mathrsfs}
\usepackage{subfigure}
\usepackage{hyperref}

\usepackage[ansinew]{inputenc}
\usepackage[pdftex]{graphicx}
\usepackage{bbm}

\def\phi{\varphi }
\def\epsilon{\varepsilon}

\swapnumbers

\theoremstyle{plain}
\newtheorem{theorem}{Theorem}[section]
\newtheorem{corollary}[theorem]{Corollary}
\newtheorem{lemma}[theorem]{Lemma}

\theoremstyle{definition}
\newtheorem{definition}[theorem]{Definition}
\newtheorem{remark}[theorem]{Remark}

\newcommand{\keywords}[1]
           {\begin{center}
            \begin{minipage}{315.83pt}
            \small
            \noindent \emph{Keywords:}~{\textrm{#1}}
            \end{minipage}
            \end{center}
            \normalsize
           }

\newcommand{\ams}[2]
           {\begin{center}
            \begin{minipage}{315.83pt}
            \small
            \noindent Classification:~Primary {\uppercase{#1}}\\
            \phantom{Classification:~}Secondary {\uppercase{#2}}
            \end{minipage}
            \end{center}
            \par\normalsize
           }           

\newcommand{\be}{\begin{equation}}
\newcommand{\ee}{\end{equation}}
\newcommand{\beo}{\begin{equation*}}
\newcommand{\eeo}{\end{equation*}}
\newcommand{\bea}{\begin{eqnarray}}
\newcommand{\eea}{\end{eqnarray}}
\newcommand{\beao}{\begin{eqnarray*}}
\newcommand{\eeao}{\end{eqnarray*}}

\newcommand{\EAB}{\mathcal{E}(p,\lambda)}
\newcommand{\Eab}{\mathcal{E}(1/2,\lambda)}
\newcommand{\IR}{\mathbb{R}}
\newcommand{\IC}{\mathbb{C}}
\newcommand{\IP}{\mathbb{P}}
\newcommand{\IN}{\mathbb{N}}
\newcommand{\IE}{\mathbb{E}}
\newcommand{\IV}{\mathbb{V}}
\newcommand{\IF}{\mathcal{F}}

\newcommand{\IT}{\mathbbm{1}_{\{N_{T-t}=n\}}}
\newcommand{\ITT}{\mathbbm{1}_{\{N_{T-t}=2n\}}}
\newcommand{\ITTT}{\mathbbm{1}_{\{N_{T-t}=2n+1\}}}
\begin{document}
\title{Continuous time Ehrenfest process in term structure modelling}
\author{
Alexander Kaplun\\
Fachbereich Mathematik, Technische Universit\"at Dortmund\\
          Vogelpothsweg 87, D-44221 Dortmund, Germany\\
e-mail: alexander.kaplun@math.tu-dortmund.de}
%\date{\today}




%\author{Alexander Kaplun}


%\institute{A. Kaplun \at
%              Technische Universität Dortmund \\
%              Tel.: +49-231-755-3064\\
%              %Fax: +123-45-678910\\
%              \email{Alexander.Kaplun@math.uni-dortmund.de}           %  \\
%%             \emph{Present address:} of F. Author  %  if needed
%%           \and
%%           S. Author \at
%%              second address
%% The correct dates will be entered by the editor
\maketitle
\begin{abstract}
In this paper, a finite-state mean-reverting model for the short-rate, based on the continuous time Ehrenfest process, 
will be examined. Two explicit pricing formulae for zero-coupon bonds will be derived in the general and the special symmetric cases. Its limiting relationship to the Vasicek model will be examined with some numerical results.%\keywords{First keyword \and Second keyword \and More}
\end{abstract}

 % insert keywords separated by a semicolon

\ams{91G30}{60J28; 91G20; 33C52} % insert the primary Maths Subject Classification number in the first bracket
         % and the secondary ams number(s) in the second bracket
         % e.g. \ams{60E20}{49G03;49F10}

\section{Introduction}
\label{intro}
One of the fundamental approaches to term structure modelling is based on the specification of 
the short-term interest rate -- the short-rate. Vasicek \cite{Ref_Vasicek} first introduced a mean-reverting 
short-rate model with Gaussian distribution and derived a closed-form representation for the zero-coupon bond (ZCB) price.  
Since then, a variety of short-rate models have become established, each having its advantages and disadvantages. 

Albeit the earliest, Vasicek's model is still very popular among practitioners owing to its analytical tractability with 
regard to ZCB prices and European options thereof. Unfortunately, the model has some shortcomings. The most prominent of these is the possibility for the interest rates to become negative -- a fact relating to all models with Gaussian distribution. Even though the probability of negative rates is rather small, not only is the realism of the model questionable, but also problems may appear while valuing ZCBs with a long time to maturity and a low interest rate level.      

The idea of using both the discrete and the continuous time versions of the Ehrenfest process in finance is well-known.
The discrete time approach was used, for example , by Okunev and Tippett \cite{Ref_Okunev} in modelling accumulated cashflows,
by Takahashi \cite{Ref_Takahashi} in exploring changes in stock prices and exchange rates for currencies, and by Buehlmann
\cite{Ref_Buehlmann} in modelling interest rates. With regard to the modelling of interest rates, it seems that the discrete time approach leads in general only to a recursively computable term structure. Sumita, Gotoh and Jin \cite{Ref_Sumita} studied the passage times and the historical maximum of the Ornstein-Uhlenbeck process via an approximation by means of a special case of the continuous time Ehrenfest process. 

This paper proposes a finite-state mean-reverting model for the short-rate related to the continuous time Ehrenfest process. By choosing arbitrary lower and upper bounds for the rate, the respective short-rate process can be seen as a suitably linearly transformed birth-and-death process on $\{0,1,\dots,N\},\ N\in \IN$. By choosing the lower bound as non-negative, the problem of negative interest rates can be avoided. Furthermore, the model allows an explicit evaluation of ZCB prices. In this way, the model aims at realism and analytical tractability. 

The main outcome of the paper is the derivation of pricing formulae for ZCBs in the general and the special symmetric cases of the model. In both cases the arbitrage-free ZCB price at time $t$ and maturity $T$ is given as follows:
\beo
 P(t,T) = C \cdot P_1(t,T)^k \cdot P_0(t,T)^{N-k},
\eeo
where $C$ is a constant, $k \in \{0,1,\dots N\},$ and $P_1$ and $P_0$ can be expressed in terms of the $\ _1F_1$ hypergeometric functions of a matrix argument given in Section \ref{sec:spec:fun} (see also \cite{Ref_Gross_2}). In the general case the model is governed by five parameters -- a valuable fact considering the fitting of the model to the market data. The special case provides four parameters and is characterized by the symmetry of the underlying distribution with respect to the mean-reverting value. The advantage here is that we have more tractable expressions of $P_1$ and $P_0$ from the computational point of view. Moreover, a suitably transformed symmetric case of the model yields the Vasicek model in the limit as $N$ tends to infinity.

The paper is organized as follows: the following section gives a short review of the special functions we shall encounter
throughout this paper. Section three deals with the Ehrenfest process in continuous time. The main results are given in 
section four, where the \textit{Ehrenfest short-rate model} is defined and the ZCB pricing formulae are derived.
The fifth section gives an overview of the Vasicek model and its limiting relationship to the Ehrenfest short-rate model. Section six illustrates the advantages of the Ehrenfest short-rate model.

This paper comprises part of my Ph.D. project at the Technische Universit\"at Dortmund. I am deeply indebted to
Professor Michael Voit for his patience and for being a wonderful research mentor.
\section{Special functions and orthogonal polynomials}
\label{sec:spec:fun}
Throughout this paper we will make use of some well-known facts concerning the \textit{Krawtchouk polynomials} (see \cite{Ref_Karlin_3}, \cite{Ref_Szego} and \cite{Ref_Voit}) and $\ _1F_1$ functions (see \cite{Ref_Gross_2}), as well as some of their practical implications. In the interests of clarity we give in this section an overview of these special functions.
%defined in the general setting as given in the following definition
%\subsection{Some facts about Krawtchouk polynomials}
%\label{sec:krawtchouk}
%Throughout this paper we will use some well-known facts on Krawtchouk polynomials, as well as some consequences. 
%In the following, we review the background formulae for the convinience of the reader.
\paragraph{Hypergeometric functions of matrix argument}

\begin{definition}
\begin{enumerate}
	\item [(a)] A \textit{partition} $m := (m_1,m_2,\dots,m_n)$ is an $n$-tuple $(n \in \IN)$ of non-negative integers such that 
	            $m_1 \geq m_2 \geq \dots \geq m_n.$
	\item [(b)] For a partition $m$ the \textit{generalized Pochhammer symbol} is defined by
	            \beo
                [a]_m := \prod_{j=1}^n (a-j+1)_{m_j},
              \eeo
              where $(a)_k := a(a+1)\cdots (a+k-1)$ denotes the usual \textit{Pochhammer symbol}. 
  \item [(c)] For a partition $m$ the \textit{normalized Schur function of index $m$} is defined by
              \beo
                Z_m(z) := |m|!\cdot \frac{\prod_{1 \leq j < k \leq n}(m_j-m_k-j+k)}{\prod_{j=1}^n (m_j+n-j)!} \cdot 
                          \frac{\det\bigl(z_i^{m_j+n-j}\bigr)}{\prod_{1 \leq i < j \leq n} (z_i - z_j)}.
              \eeo      
  \item [(d)] The \textit{hypergeometric function $ _pF_q $ of matrix argument} is defined as a real-analytic function on the                space $S_n \ (n \in \IN)$ of $n \times n$ Hermitian matrices with eigenvalues $z := (z_1,\dots,z_n)^T \in \IR^n,$                and is given by the series
              \be \label{hyp-formel-gen}
                _pF_q(a_1,\dots,a_p;b_1,\dots,b_q;z) := \sum_{j = 0}^{\infty} \frac{1}{j!} \sum_{|m| = j} 
                                             \frac{[a_1]_m\dots [a_p]_m}{[b_1]_m\dots [b_q]_m}\cdot Z_m(z),
              \ee
              where, for $1 \leq i \leq p$ and $1 \leq j \leq q,\ a_i \in \IC$ and none of the numbers $-b_i+j-1$ is a                        non-negative integer. 
\end{enumerate}
\end{definition}
%The question of convergence of $ _pF_q $ function is answerd by the following theorem adopted from \cite{Ref_Gross_1}. 
%\begin{theorem}
%\begin{enumerate}
%	\item [(a)] If $p \leq q$ then the hypergeometric series (\ref{hyp-formel-gen}) converges absolutely
%	            for all $z \in S_n.$
%	\item [(b)] If $p = q + 1$ then the series (\ref{hyp-formel-gen}) converges absolutely for $\left\| z\right\| < 1$
%	            and diverges for $\left\| z\right\| > 1.$
%	\item [(c)] If $p > q$ then the series (\ref{hyp-formel-gen}) diverges unless it terminates.            
%\end{enumerate}
%\end{theorem}
%%%%%%%%%%%%%%%%%%%%%%%%%%%%%%%%%%%%%%%%%%%%%%%%%%%%%%%%%%%%%%%%%%%%%%%%%%%%
%\begin{definition}\label{hyp_definition}
%\end{definition}
%
%$m := (m_1,m_2,\dots,m_n)$ is a $n$-tuple of nonnegative integers such that $m_1 \geq m_2 \geq \dots \geq m_n$ called %\textit{partition,} 
%for which
%is the \textit{generalized Pochhammer symbol} where
%\beo
% (a)_k := a(a+1)\dots (a+k-1)
%\eeo
%denotes the usual Pochammer symbol and $Z_m$ is the \textit{normilized Schur function of index $m$} given by
%\beo
% Z_m(z) := |m|!\cdot \frac{\prod_{1 \leq j < k \leq n}(m_j-m_k-j+k)}{\prod_{j=1}^n (m_j+n-j)!} \cdot 
%         \frac{\det\bigl(z_i^{m_j+n-j}\bigr)}{\prod_{1 \leq i < j \leq n} (z_i - z_j)}.
%\eeo
%%and
%%\beo
%%\det\bigl(z_i^{m_j+n-j}\bigr) := 
%%  \begin{vmatrix}
%%    a_{11} & a_{12} & \dots & a_{1n} \\
%%    a_{21} & a_{22} & \dots & a_{2n} \\
%%    \vdots & \vdots & \ddots & \vdots \\
%%    a_{n1} & a_{n2} & \dots & a_{nn} \\
%%  \end{vmatrix}.
%%\eeo
We will be particularly concerned with the $ _1F_1$ function, which is also known as the \textit{confluent hypergeometric function of matrix argument}. From Theorem 4.1 in \cite{Ref_Gross_2} we know that it converges absolutely for all $ z \in S_n.$ 
An important result, that will be crucial later on, is given without proof in the following remark
(see \cite{Ref_Gross_2}, p. 25). 
%The reason for our interest is that certain multiple integrals can be expressed in terms of $\ _1F_1$ functions, which will
%be crutional later on.
%the possibility of expressing certain multiple integrals,
%that have to be computed iteratively, by $_1F_1$ function. This result is given without proof in the following remark
%(see \cite{Ref_Gross_2}, p. 25).

\begin{remark}
Let $\Delta_n$ denote the \textit{standard simplex} in $\IR^n\ (n\in \IN),$ defined by
\be\label{def-simplex}
 \Delta_n := \left\{(x_1,\dots,x_n)\in \IR^n :x_i \geq 0, i=1,\dots n, \sum_{i=1}^n x_i \leq 1 \right\}.
\ee 
Then for $a > 0$ the following equation holds:
\be \label{simplex-trafo-allg}
  _1F_1(1;a+n;z) = (a)_n \int_{\Delta_n} \left(1-\sum_{i = 1}^n \right)^{a-1} \cdot
                   \exp\left(\sum_{i=1}^n z_i x_i\right)\, dx_1\dots dx_n .
\ee
\end{remark}
In order to compute the $ _pF_q$ function numerically we truncate the series (\ref{hyp-formel-gen}) by $m \leq H$ 
as follows:
\be\label{hyp-trunvated}
 _pF_q^H(a_1,\dots,a_p;b_1,\dots,b_q;z) := \sum_{j = 0}^{H} \frac{1}{j!} \sum_{|m| = j} 
                                             \frac{[a_1]_m\dots [a_p]_m}{[b_1]_m\dots [b_q]_m}\cdot Z_m(z).
\ee 
Koev and Edelman \cite{Ref_Koev} provide an effective algorithm for computing the $ _pF_q^H$ function. For $ z \in S_n$
the complexity of their algorithm is linear in $n$ and subexponential in $H,$ which is acceptable
if we consider the fast convergence of the power series (\ref{hyp-trunvated}).
%The numerical computation of more than complicated $ _pF_q$ functions will be discussed later on in Section \ref{sec:numeric}. 
%, that can be easely computed numerically (see Section \ref{sec:numeric}). This technics extend the results in \cite{Ref_Delbaen} in a similar way. 
%%%%%%%%%%%%%%%%%%%%%%%%%%%%%%%%%%%%%%%%%%%%%%%%%%%%%%%%%%%%%%%%%%%%%%%%%%%%%%%%%%%%%%%%%%%%%%%%%%%%%%%%%%%%%%%%%%%%%%%5
\paragraph{Krawtchouk polynomials}
For given $N\in \IN$ and $0 < p = 1 - q < 1$ the Krawtchouk polynomials $K_l(x) := K_l(x;N;p)$ are the 
orthogonal polynomials that relate to the binomial distribution $B_{N,p}$ and the probability mass function
$ \omega(x) := \binom{N}{x}p^xq^{N-x}$ at the points $x = 0,1,\dots,N.$
They can be defined in two different, but equivalent ways.
\begin{definition} \label{krawtchouk-definition}
For $x,l \in \{0,1,\dots,N\}$ we set 
\be \label{krawtchouk-definition1}
 K_l(x) :=\, _2\mathcal{F}_1(-l,-x;-N;1/p) := \sum_{k=0}^N \frac{(-l)_k (-x)_k}{(-N)_k k!} \left(\frac{1}{p}\right)^k,
\ee 
or
\be \label{krawtchouk-definition2}
 K_l(x) := \binom{N}{l}^{-1} \sum_{k=0}^N (-1)^k \binom{N-x}{l-k} \binom{x}{k} \left(\frac{q}{p}\right)^k,
\ee 
where $\, _2\mathcal{F}_1$ is the classical Gauss hypergeometric function (see \cite{Ref_Szego}, §4.21). 
\end{definition}
Definition \ref{krawtchouk-definition} leads to the following basic well-known properties:
\begin{lemma} \label{k-lemma}
\begin{enumerate}
	\item [$(a)$] Symmetry:
	   \be \label{k_symmetry}
	       K_l(x) = K_x(l)
	   \ee
	   for all $x,l \in \{0,1,\dots,N\}$.
	\item [$(b)$] $K_0(x) = K_l(0) = 1$ for all $x,l \in \{0,1,\dots,N\}$.
	\item [$(c)$] $K_1(x) = 1-\frac{x}{Np}.$ 
	\item [$(d)$] Generating function:
	  \be \label{k_generating}
         \bigl(1- \frac{q}{p}\cdot s\bigr)^i \cdot \bigl(1+ s\bigr)^{N-i} = \sum_{k = 0}^{N} \binom{N}{l}K_{l}(i)s^l  
    \ee
    for all $x\in \{0,1,\dots,N\}.$
  \item [$(e)$] Recurrence relation:
    \be \label{k_recurrence}
        -xK_l(x) = (N-l)pK_{l+1}(x) - \left[(N-l)p + lq\right]K_l(x) + lqK_{l-1}(x) 
    \ee 
    for all $x,l \in \{0,1,\dots,N\}.$
  \item [$(f)$] Orthogonality relation:
	  \be \label{k_orthogonality}
         \sum_{x = 0}^N  K_l(x) K_m(x) \omega(x)= \frac{\delta_{l,m}}{\pi_l},  
    \ee      
    where 
    \be \label{k_relation}
       \pi_l := \binom{N}{l}\left(\frac{p}{q}\right)^l = \omega(l)q^{-N}
    \ee
    for all $l,m\in \{0,1,\dots,N\}.$
  \item [$(g)$] For $l,m \in \{0,1,\dots,N\},\ m \leq l$
	  \be \label{k_ortho-recurrence}
         B_{m,l} := \sum_{x = 0}^N  x K_l(x) K_m(x)\omega(x) = 
             \begin{cases}
                \phantom{-}0 					 \quad & \text{if}\quad m \leq l-2, \\
                -lq \pi_{l-1}					 \quad & \text{if}\quad m = l-1,\\
                \phantom{-}((N-l)p + lq)\pi_l \quad & \text{if}\quad m = l.
             \end{cases}  
    \ee  
    Note that $B_{l,m} = B_{m,l}.$    
\end{enumerate}
\end{lemma}
\proof
$(a)-(e)$ follow from (\ref{krawtchouk-definition1}), $(f)$ follows from (\ref{krawtchouk-definition2}) and
$(g)$ is a direct application of (\ref{k_recurrence}) and (\ref{k_orthogonality}) (see e.g. \cite{Ref_Szego}, §2.82).
%, \cite{Ref_Voit} or \cite{Ref_Karlin_3} for more details.  
\qed
%%%%%%%%%%%%%%%%%%%%%%%%%%%%%%%%%%%%%%%%%%%%%%%%%%%%%%%%%%%%%%%%%%%%%%%%%%%%%%%%%%%%%%%%%%%%%%%%%%%%%%%%%%%%%%%%%%%%%%%%%%
\section{Original Ehrenfest model} \label{sec:e-model}
The original Ehrenfest model describes the heat exchange between two isolated bodies, each of arbitrary temperature. The temperatures are symbolized by the number of fluctuating balls in two urns with a total of $N \in \IN$ balls. 
For details of the continuous and discrete time versions of the model we refer to \cite{Ref_Bingham}, \cite{Ref_Karlin_3}, \cite{Ref_Kraft} and \cite{Ref_Siegert}.
%In this model $N \in \IN$ balls are
%distributed initially between two urns where the number of balls in each urn symbolizes the bodies' temperature.

In this section we shall discuss the continuous time Ehrenfest process. Primarily, its representation as a sum of independent simple processes will allow us to show the main result of this paper which comes up in the next section. Furthermore, we examine the transition semigroup of the Ehrenfest process and explore some of its basic properties. 
%Originally, considered in discrete time, the model was given by a Markov chain $(X_n)_{n\in\IN}$ on the state space 
%$\{0,1,\dots,N\}$ denoting the number of balls in one of the urns.
%However, it can be formulated as simple ball and urn model; the balls correspond to the molecules and the urns to the two %containers\footnote{Internetseite}.
%The represantation of the transition probabilities will involve some orthogonal polynomials, named Krawtchouk polynomials, %which definition and properties we will give at first in this section.  
\paragraph{Ehrenfest process}
\label{subsec:E-process}
Let $N$ balls, initially distributed between urns I and II, fluctuate independently in continuous time between the two urns. We fix a fluctuation parameter $\lambda > 0$ and independent Poisson processes $(N^{1}_t)_{t\geq 0},\dots,(N^{N}_t)_{t\geq 0}$ with intensity $\lambda.$ Let $(\hat{Y}_n)_{n\in \IN}$ be a Markov chain with the state space $\{0,1\}$ and transition probability matrix
\be \label{k-matrix}
 P := \begin{pmatrix} 1-\alpha & \alpha \\
                     \beta & 1-\beta \\
 \end{pmatrix}.
\ee 
%i.e. $(\hat{Y}_n)_{n\in \IN}$ is a two-parameter Ehrenfest chain with only one ball.  
Then, the subordinated Markov chain $(Y^{l}_t := \hat{Y}_{N^{l}_t})_{t \geq 0}$  describes the state of the $l$-th ball at time $t,$ where $Y^l_t = 1$ or $0$ when the $l$-th ball is in urn I or II respectively.
%\beo
%  Y^l_t = \begin{cases}
%             1 \quad & \text{if $l$-th ball is in urn I},\\
%             0 \quad & \text{if $l$-th ball is in urn II}.
% \end{cases}
%\eeo   
Hence,  
\be \label{substitution-formula}
 \Bigl(X_t := \sum_{l = 1}^{N} Y^l_t\Bigr)_{t\geq 0}
\ee 
is a Markov process with the state space $E := \{0,1,\dots,N\},$ denoting the number of balls in urn I at time $t.$ 
We call $(X_t)_{t \geq 0}$ the \textit{(continuous time) Ehrenfest process}. A discrete time version of (\ref{substitution-formula}) with arbitrary $\alpha$ and $\beta$ was studied by Kraft and Schaefer \cite{Ref_Kraft}.
%Thus, $(X_t)_{t \geq 0}$ is a \textit{birth-and-death process}, which discrete time analog, known as a 
%\textit{two-parameter Ehrenfest chain,} was discussed for instance in \cite{Ref_Kraft}. 
\begin{remark}\label{bd-ehrenfest}
\begin{enumerate}
	\item [(a)] A special case of (\ref{substitution-formula}) with $\alpha = \beta = 1,$ first suggested by Siegert                            \cite{Ref_Siegert} and also studied by Bingham \cite{Ref_Bingham}, where the transitions become
	             ``deterministic'' in the sense of switching between the states 0 and 1, will be important for us later on in
	             Section \ref{sec:2}. (Its discrete time analogue leads to the original Ehrenfest chain).  
  \item [(b)] Karlin and McGregor \cite{Ref_Karlin_3} provided an alternative but equivalent definition of 
              (\ref{substitution-formula}) as a \textit{birth-and-death process} with the state space $E$. Here, the time                     intervals between events are independently exponentially distributed with intensity $\gamma,$ and for $i \in E$                 the birth and death rates are $\lambda_i := \gamma \alpha \frac{(N-i)}{N}$ and $\mu_i = \gamma \beta \frac{i}{N}$               respectively, where $\alpha$ and $\beta$ are given as above. It can be verified that in the setting at hand                     $\gamma = \lambda N.$           
\end{enumerate}
\end{remark}%Originally, the Ehrenfest model was considered in discrete time. Here, a ball is selected at random each having
%probability $\frac{1}{N}.$ If the selected ball is in urn I, it changes the urn with probability $0 < \alpha \leq 1$
%and stays in urn I with probability $1-\alpha.$ If the selected ball is in urn II, it changes the urn with probability
%$0 < \beta \leq 1$ and stays in urn II with probability $1-\beta.$
%%(case $\alpha = \beta = 1$ leads to the original Ehrenfest chain). 
%The process is repeated any number of times. Then, the number of balls in urn I after $n$ selections, denoted by
%$(\hat{X}_n)_{n\in \IN},$ is a Markov chain with transition probability matrix
%%$P = (p_{ij}),$ where% $(p_{ij})_{i,j = 1}^n$
%\be\label{e-transition}
% p_{ij} = \begin{cases}
%              \alpha\cdot\frac{N-i}{N} \quad & \text{if}\quad j = i + 1,\\
%              \beta\cdot\frac{i}{N}		\quad & \text{if}\quad j = i - 1,\\
%              1 - \alpha\cdot\frac{N-i}{N} - \beta\cdot\frac{i}{N} \quad & \text{if}\quad j = i, \\
%              0 										\quad & \text{otherwise,}
%           \end{cases}.
%\ee
%which is known as a \textit{two-parameter Ehrenfest chain} first introdused in 1993 by Kraft and Schaefer \cite{Ref_Kraft},
%who provided following result.
%% We will exploit this approach  
 
%This approach will provide us more applicable results. 

Before computing the transition semigroup of $(X_t)_{t \geq 0}$ we need the following result:
\begin{lemma} \label{binaryprocess-transprob}
The transition semigroup of $(Y^{l}_t)_{t \geq 0}$ is given by
\be\label{y_semigroup}
 P(t) = 
        \begin{pmatrix} q + pe^{-\lambda (\alpha + \beta)t}\ & p - pe^{-\lambda (\alpha + \beta)t} \\
                 q - qe^{-\lambda (\alpha + \beta)t}\ & p + qe^{-\lambda (\alpha + \beta)t} \\ 
        \end{pmatrix},
\ee
where $p := \frac{\alpha}{\alpha+\beta}$ and $q := 1-p.$
\end{lemma}
\proof
Since $(Y^{l}_t)_{t \geq 0}$ is a Markov chain subordinated by a Poisson process with index $\lambda,$ the associated
transition semigroup can be written as (see \cite{Ref_Bremaud}, p.333)
\be \label{trans-prob}
 P(t) = e^{-\lambda t}\cdot e^{\lambda t P} := e^{-\lambda t} \sum_{n=0}^{\infty} \frac{(\lambda t)^n}{n!} P^{n}.
\ee
We can avoid the computation of $P^n$ by writing %Let $\alpha + \beta \leq 1.$ 
$P = \mu S + (1-\mu)I,$ where 
      $S = \begin{pmatrix} q\ & p \\
                           q\ & p \\ 
           \end{pmatrix}$ 
is a stochastic matrix with $S^2 = S, \ I$ is the identity matrix, 
and $\mu := \alpha + \beta.$
Then, (\ref{trans-prob}) becomes 
\beao
 P(t) &=& e^{-\lambda t}\cdot e^{\lambda t (\mu S + (1-\mu)I)} 
       = e^{-\lambda t}\cdot e^{\lambda\mu t S } \cdot e^{\lambda(1-\mu) t I } \\
      &=& e^{-\lambda\mu t}\cdot e^{\lambda\mu t S}
       = e^{-\lambda (\alpha + \beta) t} \sum_{n=0}^{\infty} \frac{(\lambda (\alpha + \beta) t)^n}{n!} S^{n}.
\eeao
Since $S^n = S$ for all $n \in \IN,$ we can easily compute the above series, which completes the proof.
%\beo
%P(t) = e^{-\lambda (\alpha + \beta) t} \sum_{n=0}^{\infty} \frac{(\lambda (\alpha + \beta) t)^n}{n!} S^{n}.
%\eeo  
%In case of $\alpha + \beta > 1,$ we obtain the same result by writing $S = \mu P + (1-\mu)I$
%if we set $\mu := \frac{1}{\alpha + \beta}.$ Since $S^n = S$ for all $n \in \IN,$ we can easely compute
%the above series, completing the proof.
\qed  
\begin{remark}\label{dependance}
Analogue to the proof above, we see that a variation of $\alpha$ and $\beta$ in (\ref{y_semigroup}) can be equivalently
described as a suitable modification of $p$ and $\lambda$. Thus, the distribution of $(Y_t)_{t\geq 0}$ depends only on 
the two parameters $p$ and $\lambda$ or, equivalently, $\alpha$ and $\beta$.
\end{remark}

\begin{theorem}(Properties of the Ehrenfest process)
\label{trm-e-process}
Let $(X_t)_{t\geq 0}$ be the Ehrenfest process given by (\ref{substitution-formula}).
\begin{enumerate}
	\item [$(a)$] The transition probabilities $p_{ij}(t) := \IP(X_{t+s}=j | X_s = i)$ are given by
	  \be \label{e-trans-prob}
	    p_{ij}(t) = \binom{N}{j} \left(\frac{p}{q} \right)^j \sum_{x = 0}^{N} \binom{N}{x} p^x q^{N-x}
	               K_i(x) K_j(x) e^{-\lambda (\alpha+\beta)xt},
	  \ee
%	  where $p, q$ and $K_i(.),\ i=0,\dots,N,$ are given as in Theorem \ref{trm-e-chain}.
    where  $p := \frac{\alpha}{\alpha+\beta},\ q := 1 - p,$ and $K_i(.) := K_i(.;p;N)\ (i \in E)$ are the Krawtchouk 
	  polynomials as given in Definition \ref{krawtchouk-definition} .
	\item [$(b)$] The conditional mean and variance of $(X_t)_{t\geq 0}$ are given by 
	   \bea\label{e_mean}
	     \IE[X_t | X_0 = i] &=& Np - (Np-i)e^{-\lambda(\alpha+\beta)t}, \\
	   \label{e_variance}
	     \IV ar[X_t | X_0 = i] &=& Np(1-p) + (Np-i)(2p-1)e^{-\lambda(\alpha+\beta)t} \\
	         && -(Np-i)^2(2p-1)e^{-2\lambda(\alpha+\beta)t}. \nonumber
	   \eea  
	\item [$(c)$] The stationary distribution is the limiting distribution, and it is given by the binomial distribution                          $B_{N,p}$ on $E$ with parameter $p.$ 
\end{enumerate}
\end{theorem}
\proof 
Ad $(a):$ The proof here is similar to that in \cite{Ref_Bingham}. First, we compute the moment-generating function 
of $(Y^{l}_t)_{t \geq 0},$ given $Y^{l}_0.$ In the following we suppress the dependence of $(Y^{l}_t)_{t \geq 0}$ on $l$ 
when it is clear from the context. From Lemma \ref{binaryprocess-transprob} we have 
\beao
\IE \left[z^{Y_t} \big| Y_0 = 1\right] &=& p_{10}(t) + p_{11}(t)z = q + pz - q(1-z)e^{-\lambda (\alpha + \beta)t},\\
\IE \left[z^{Y_t} \big| Y_0 = 0\right] &=& p_{00}(t) + p_{01}(t)z = q + pz + p(1-z)e^{-\lambda (\alpha + \beta)t}.
\eeao
Since $(Y^{l}_t)_{t \geq 0}$ are independent %and have the same law 
for all  $l = 1,\dots,N,$ we obtain from (\ref{substitution-formula}) the moment-generating function
of $(X_t)_{t\geq 0},$ given $X_0 = i,$ as follows:  
\beao
\sum_{j = 0}^{N} p_{ij}(t)z^j &=& %\IE \left[z^{X_t} \big| X_0 = i\right] =
                                  \IE \left[z^{\sum_{l = 1}^{N} Y^l_t} \big| X_0 = i\right] 
                                = \IE \left[\prod_{l = 1}^{N} z^{Y^l_t} \big| X_0 = i\right] \\
 &=& \IE \left[z^{Y_t} \big| Y_0 = 1\right]^{i}\cdot \IE \left[z^{Y_t} \big| Y_0 = 0\right]^{N-i} \\
 &=& \left[q + pz - q(1-z)e^{-\lambda (\alpha + \beta)t}\right]^{i} \cdot
     \left[q + pz + p(1-z)e^{-\lambda (\alpha + \beta)t}\right]^{N-i} \\
 %&=& (q+pz)^{N} \cdot \left[1 - q\frac{(1-z)}{q+pz}e^{-\lambda (\alpha + \beta)t}\right]^{i} \cdot
 %                      \left[1 + p\frac{(1-z)}{q+pz}e^{-\lambda (\alpha + \beta)t}\right]^{N-i} \\
 &=&   (q+pz)^{N} \cdot \left[1 - \frac{q}{p}\cdot \frac{p(1-z)}{q+pz}e^{-\lambda (\alpha + \beta)t}\right]^{i} \cdot
                       \left[1 + \frac{p(1-z)}{q+pz}e^{-\lambda (\alpha + \beta)t}\right]^{N-i}.                               \eeao
%Applying $(3.3)$ from \cite{Ref_Karlin_3}, where the generating function of the Krawtchouk polynomials is given by
%\be 
% (1- \frac{q}{p}\cdot s)^i \cdot (1+ s)^{N-i} = \sum_{k = 0}^{N} \binom{N}{k}K_{k}(i)s^k  
%\ee                    
%for $i = 0,1,\dots,N$ and $K_i(k) = K_i(k;p,N),$ and using the symmetry relation 
%$K_i(k) = K_k(i),\ i,k = 0,\cdots, N,$
Applying $(a)$ and $(d)$ of Lemma \ref{k-lemma}, we obtain %with $s = \frac{p(1-z)}{q+pz}e^{-\lambda (\alpha + \beta)t}$
\beao
\sum_{j = 0}^{N} p_{ij}(t)z^j 
  &=& (q+pz)^{N} \sum_{x = 0}^{N} \binom{N}{x}K_{i}(x) \left(\frac{p(1-z)}{q+pz}e^{-\lambda (\alpha + \beta)t}\right)^x \\
  &=& \sum_{x = 0}^{N} \binom{N}{x}K_{i}(x)(q+pz)^{N-x} p^x (1-z)^x e^{-\lambda (\alpha + \beta)xt} \\
  &=& \sum_{x = 0}^{N} \binom{N}{x}p^x q^{N-x}K_{i}(x) 
      \left(1 - \frac{q}{p}\cdot\frac{pz}{q}\right)^x\left(1 + \frac{pz}{q}\right)^{N-x} e^{-\lambda (\alpha + \beta)xt}.
\eeao
Applying (\ref{k_generating}) once again, we get
\beao
\sum_{j = 0}^{N} p_{ij}(t)z^j 
  &=& \sum_{x = 0}^{N} \binom{N}{x}p^x q^{N-x}K_{i}(x) 
      \left[\sum_{j = 0}^{N} \binom{N}{j} K_{j}(x)\left(\frac{pz}{q}\right)^j\right]e^{-\lambda (\alpha + \beta)xt} \\
  &=& \sum_{j = 0}^{N} \left[\binom{N}{j} \left(\frac{p}{q}\right)^j
      \sum_{x = 0}^{N} \binom{N}{x}p^x q^{N-x} K_{i}(k)K_{j}(x)e^{-\lambda (\alpha + \beta)xt}\right]z^j.   
\eeao 
Equating coefficients of $z^j$ completes the proof of the claim.\\
Ad $(b):$ Combining (\ref{e-trans-prob}) with results $(a), (b)$ and $(g)$ of Lemma \ref{k-lemma}, we obtain
\bea \label{e-calc1}
  \IE[X_t | X_0 = i] &=& \sum_{j=0}^N j\cdot p_{ij}(t) 
  = \sum_{j=0}^N j \cdot\omega(j) \sum_{x = 0}^{N} \pi_x K_i(x) K_j(x) e^{-\lambda(\alpha+\beta)xt} \nonumber\\
  &=& \sum_{x = 0}^{N} \pi_x K_i(x)e^{-\lambda(\alpha+\beta)xt} \sum_{j=0}^N j\cdot K_x(j) \omega(j) \nonumber\\
  &=& \sum_{x = 0}^{N} \pi_x B_{0,x} K_i(x) e^{-\lambda(\alpha+\beta)xt}, 
\eea
where $B_{0,x} \neq 0$ for $x \in \{0,1\}$ as given in (\ref{k_ortho-recurrence}), which implies (\ref{e_mean}).
Moreover,
\beao 
  \IE[X^2_t | X_0 = i] &=& \sum_{j=0}^N j^2\cdot p_{ij}(t) 
  = \sum_{j=0}^N j^2 \cdot\omega(j) \sum_{x = 0}^{N} \pi_x K_i(x) K_j(x) e^{-\lambda(\alpha+\beta)xt} \\
  &=& \sum_{x = 0}^{N} \pi_x K_i(x)e^{-\lambda(\alpha+\beta)xt} \sum_{j=0}^N j^2\cdot K_x(j) \omega(j) \\
  &=& \sum_{x = 0}^{N} \pi_x K_i(x)e^{-\lambda(\alpha+\beta)xt} \sum_{j=0}^N j\cdot\omega(j) 
      \Bigl[-(N-x)pK_{x+1}(j) \\
  &&\qquad\qquad\qquad \quad\     + \left[(N-x)p+xq \right]K_x(j)-xqK_{x-1}(j)  \Bigr] \\
  &=& \sum_{x = 0}^{N} \pi_x K_i(x)e^{-\lambda(\alpha+\beta)xt} 
   \Bigl[-(N-x)pB_{0,x+1} \\
  &&\qquad\qquad\qquad \qquad\ \   + \left[(N-x)p+xq \right]B_{0,x} -xqB_{0,x-1} \Bigr], 
 % &=& \sum_{x = 0}^{N} \pi_x B_{0,x} K_i(x) e^{-\lambda(\alpha+\beta)xt}, 
\eeao  
where $B_{0,x}$ is given by (\ref{k_ortho-recurrence}). A straightforward computation of 
\beo
 \IV ar[X_t | X_0 = i] = \IE[X^2_t | X_0 = i] - \IE[X_t | X_0 = i]^2
\eeo
leads immediately to (\ref{e_variance}).\\
Ad $(c):$ Clear.
\qed   
%Motivated by the fact that $(X_t)_{t\geq 0}$ and $(\hat{X}_n)_{n\in \IN}$ discribe the number of the balls
%in urn I at time $t$ in continouos time and after $n$ transitions in discrete time respectively, we conclude
%the following result. 
%\begin{corollary}
%Let $(X_t)_{t\geq 0}$ be given by (\ref{substitution-formula}), $(\hat{X}_n)_{n\in \IN}$ 
%be given by (\ref{e-transition}) %with $a = b := N\lambda$
%and $(\tilde{N}_t)_{t\geq 0}$ be an indipendent
%Poisson process with intensity $s := 2\lambda N.$ Then, $(X_t)_{t\geq 0}$ and $(\hat{X}_{\tilde{N}_t})_{t\geq 0}$
%are equal in law.
%\end{corollary}
%\proof The statement follows directly from Theorem \ref{trm-e-chain} and the representation (\ref{substitution-formula})
%of the transition semigroup $P(t)$ of the subordinated Ehrenfest chain $(\hat{X}_{\tilde{N}_t})_{t\geq 0}$. By a 
%straightforward calculation of (\ref{trans-prob}) we obtain exactly the transition probability of $(X_t)_{t\geq 0}$ given in %Theorem \ref{trm-e-process}.
%\qed
%%%%%%%%%%%%%%%%%%%%%%%%%%%%%%%%%%%%%%%%%%%%%%%%%%%%%%%%%%%%%%%%%%%%%%%%%%%%%%%%%%%%%%%%%%%%%%%%%%%%%%%%%%%%%%%%%%%%%%%%%%
\section{Ehrenfest short-rate model}
\label{sec:2}
%A modern reference on interest rate modelling can be found, for instance, in \cite{Ref_Brigo}.
In this section we introduce a finite-state mean-reverting short-rate model associated with the continuous time Ehrenfest process (\ref{substitution-formula}) and give its basic properties. As a main result, we exploit the algebraic-combinatorial roots of the Ehrenfest process and derive explicit pricing formulae for ZCBs in the general and the special cases of the process, both of which have their advantages.  

\paragraph{Definition and properties}
\label{sec:2-1}
Let $[r_m,r_M] \subseteq \IR$ be an interval on the real line. We decompose it into $N$ equal pieces of length
$h := \frac{r_M-r_m}{N}$ and consider the process 
\be \label{zinsprocess:1}
(R^{(N)}_t := h X_t^{(N)} + r_m)_{t \geq 0}
\ee
as a \textit{short-rate process} with state space $E := \{r_k := hk + r_m,\ k = 0,\dots,N\},$ where $(X_t^{(N)})_{t\geq 0}$
is the Ehrenfest process given by (\ref{substitution-formula}) with $\alpha, \beta \in (0,1]$.
Considering Remark \ref{bd-ehrenfest} $(b)$, we notice that $(R^{(N)}_t)_{t\geq 0}$ can be seen as an affine linearly transformed birth-and-death process on $\{0,1,\cdots,N\}.$ In the case at hand, $N$ can be interpreted as the state space discretization parameter. Clearly, $(R_t := R^{(N)}_t)_{t\geq 0}$ also depends on $N$. We will suppress this dependence when it is clear from the context. Bearing in mind Remark \ref{dependance}, we denote this short-rate model as $\EAB$ \textit{model}.

From Theorem \ref{trm-e-process} we immediately obtain the conditional mean and variance of $(R_t)_{t\geq 0},$ given
$R_0 := r_k \in E,$ as follows:
\bea\label{e_short_mean}
  \IE[R_t | R_0 = hk + r_m] &=& h \cdot \IE[X_t | X_0 = k] + r_m \nonumber\\
                            &=& (r_M-r_m)\Bigl(p - \bigl(p-\frac{i}{N}\bigr)e^{-\lambda(\alpha+\beta)t}\Bigr) + r_m,\\
%\eea
%\bea
\label{e_short_variance}	
	\IV ar[R_t | R_0 = hk + r_m] &=& h^2\cdot \IV ar[X_t | X_0 = k] \nonumber\\
	                             &=& \frac{(r_M-r_m)^2}{N^2}\Bigl(Np(1-p) + (Np-i)(2p-1)e^{-\lambda(\alpha+\beta)t} \nonumber\\
	&&\qquad  \qquad \quad   - (Np-i)^2(2p-1)e^{-2\lambda(\alpha+\beta)t}\Bigr), 
\eea
where $p = \frac{\alpha}{\alpha + \beta}.$ We also obtain the \textit{mean reversion} of $(R_t)_{t\geq 0}:$ 
\bea
\label{mean-rev1}
 \lim_{t \rightarrow \infty} \IE [R_t]&=& %h\cdot\lim_{t \rightarrow \infty} \IE  [X_t] + r_m 
   %= \frac{r_M-r_m}{N}Np + r_m 
   p r_M + (1-p) r_m, \\
\label{mean-rev2}   
  \lim_{t \rightarrow \infty} \mathbb{V}ar [R_t]&=& 
  %h^2\cdot\lim_{t \rightarrow \infty}\mathbb{V}ar[X_t]  
  \frac{(r_M-r_m)^2}{N}p(1-p) < \infty.  
\eea
Thus, we have a total of five parameters, $r_m,r_M,p,\lambda$ and $N,$ to fit the model to the market data.
Here, $p$ governs the skewness of the underlying distribution, $r_M-r_m, N$ and $p$ have an impact on its kurtosis, and $\lambda$ influences the speed of reversion to the mean reverting value $p r_M + (1-p) r_m.$ 
%\begin{Theorem}[Properties of Ehrenfest short-rate model]
%Thus, the mean-reverting value lies in the middle of $[a,b].$
%%%%%%%%%%%%%%%%%%%%%%%%%%%%%%%%%%%%%%%%%%%%%%%%%%%%%%%%%%%%%%%%%%%%%%%%%%%%%%%%%%%%%%%%%%%%%%%%%%%%%%%%%%%%%%%%%%%%%%%%%%%%55
\paragraph{Zero-coupon bond}
\label{subsec:zb}
We assume that an equivalent martingale measure (or risk-neutral measure) exists (see \cite{Ref_Lakner}, Prop. 4.2),
and that the underlying probability measure $\IP$ is this equivalent martingale measure.
Let $(\IF_t)_{t\geq 0}$ be the natural filtration of $(R_t)_{t\geq 0}.$ Then, the arbitrage-free ZCB price at time $t$ with a face value of 1 monetary unit and maturity at $T$ is given by (see \cite{Ref_Brigo}, p. 51)
\be\label{zb-price}
 P^{(N)}(t,T) = \IE\left[\exp\left(-\int_t^T R_s\, ds \right) \Bigg| \IF_t \right], \quad r \in E.
\ee 
In the following we also omit explicitly writing out the dependence of $P(t,T)$ on $N$ when it is clear from the context.
%The calculation of (\ref{zb-price}) associated with the two-parameter Ehrenfest process general case where $\alpha$ and %$\beta$ in (\ref{k-matrix}) are arbitrary in $[0,1]$
%First, we consider the two-parameter , where $\alpha$ and $\beta$ in (\ref{k-matrix}) are arbitrary in $[0,1]$ 
%and hence $p \in [0,1].$ We will use some results on the spectral represantation of the transition probabilities
%of birth-death processes developed in \cite{Ref_Karlin_3}. This approach will lead to an explicit formula for a zero-coupon %bond, which numerical computation time grows exponentially as the required accuracy??. Then, we restrict our model
%to the case $\alpha=\beta = 1,$ which yields a \textit{one-parameter Ehrenfest model} with $p = \frac{1}{2}.$ The crux
%here will be deterministic switching between the states and properties of zwischenwartezeiten of Poisson process.
%This approch yields a more tractable formula for a zero-coupon bond and as we will see in section (\ref{e-v-convergence}),
%the one-parameter Ehrenfest model is still very well suited for modelling term structure. 

The calculation of (\ref{zb-price}) within the $\EAB$ model with arbitrary $\alpha, \beta \in (0,1]$
is inspired by the proof of Theorem 3.1 in \cite{Ref_Delbaen}. There, Delbaen and Shirakawa represent the transition probabilities of the underlying short-rate process as a weighted series of the Jacobi polynomials. Using orthogonality relations of the Jacobi polynomials, they obtain a pricing formula for ZCBs in the associated model. However, this formula is only semi-explicit, since it contains multiple integrals that have to be calculated iteratively. We will avoid this problem by representing such integrals in terms of $ _1F_1$ functions.  
%exploited the spectral representation of the transition probabilities of the underlying short-rate process as a weighted %series of Jacobi polynomials and used thier orthogonal relations to obtain a formula for a ZCB in the associated model. %However, this formula is only semi-explicit, since it contains multiple integrals in the form of (\ref{zb-integral2}) that %have to be calculated iteratively. We will avoid this problem by representing such integrals in terms of $ _1F_1$ function. %defined in the general setting as given in the following definition.
%%%%%%%%%%%%%%%%%%%%%%%%%%%%%%%%%%%% %Theorem %%%%%%%%%%%%%%%%%%%%%%%%%%%%%%%%%%%%%%%%%%%%%%%%%%%%%%%%%%%%%%%%%
\begin{theorem}[ZCB price in $\EAB$ model] \label{trm:zbformel:p}
Let $(R_t)_{t \geq 0}$ be given by the definition (\ref{zinsprocess:1}) with $\alpha, \beta \in (0,1], 
\ p = \frac{\alpha}{\alpha+\beta}$ and $\lambda > 0.$ The price at time $t\geq 0$ of a ZCB with maturity at $T,$ conditional on $R_t = r \in E,$ is given by
\be \label{zb:ehrenfest:p}
P(t,T) = e^{-r_m(T-t)}\cdot P_1(t,T)^k \cdot P_0(t,T)^{N-k},
\ee
where $k := \frac{r - r_m}{h} \in \{0,\dots,N\}$ and for $m \in \{0,1\}$ 
\bea\label{zb-series1}
P_m(t,T) &:=& 1 + \sum_{n=1}^{\infty} (-1)^n \frac{(-h(T-t))^n}{n!} \cdot \\
         &&\sum_{i_1 = 0}^1 \cdots \sum_{i_n = 0}^1 K_m(i_n)
            \left[\prod_{j =1}^n \left(\frac{p}{q}\right)^{i_j}B_{i_{j-1},i_j}\right]  \cdot
            \, _1F_1\left(1;n+1;z^{(n)}\right). \nonumber          
\eea 
Here, $K_m$ is given according to $(b)$ and $(c)$ of Lemma \ref{k-lemma}, $\, _1F_1$ is defined by (\ref{hyp-formel-gen}),
$z^{(n)} := -\lambda(\alpha+\beta)(T-t)(i_1,\dots,i_n)^T \in \IR^{n},\ i_0 := 0,$ and 
\beo
 B_{i_{j-1},i_j} := \begin{cases}
                      i_j(p-1) \quad & \text{if}\quad i_{j-1} = i_j-1,\\
                      i_j(1-2p) + p \quad & \text{if}\quad i_{j-1} = i_j.
                   \end{cases}
\eeo
\end{theorem}
\proof
Let $r_k = hk + r_m, \ k \in \{0,\dots,N\},$ be the state of $(R_t)_{t\geq 0}$ at time $t.$ Bearing in mind that 
$(R_t)_{t\geq 0}$ is a Markov process, and using the definitions (\ref{substitution-formula}) and (\ref{zinsprocess:1}), 
we get from (\ref{zb-price})
\bea \label{zb-formel:1}
P(t,T) %&=& \IE\left[\exp\left(-\int_t^T R_s\, ds \right) \Bigg| R_t = r\right] \nonumber \\
 &=& \IE\left[\exp\left(-\int_t^T \left( h\sum_{l=1}^{N} Y^l_s + r_m \right)\, ds \right) \Bigg| X_t = k\right]\nonumber\\
 &=& e^{-r_m(T-t)}\cdot\IE\left[\prod_{l=1}^{N} \exp\left(-\int_t^T h Y^l_s \, ds \right) \Bigg| X_t = k\right] \nonumber\\
 &=& e^{-r_m(T-t)}\cdot\IE_{1,t}\left[\exp\left(-\int_t^T h Y_s \, ds \right)\right]^k
               \cdot\IE_{0,t}\left[\exp\left(-\int_t^T h Y_s \, ds \right)\right]_,^{N-k} 
 %&=& e^{-r_m(T-t)}\cdot \tilde{P}^{(N)}_1(t,T)^{k} \cdot \tilde{P}^{(N)}_0(t,T)^{N-k},             
\eea
where $\IE_{m,t}[\ .\ ] := \IE[\ .\ |Y_t = m]$ for $m \in \{0,1\}.$
The last equality holds because of the independence 
%and identical distribution 
of $(Y^l_t)_{t\geq 0}$ for all $l = 1,\dots,N.$ In the following we omit writing out the dependence on particular $l,$ and set
\be \label{zb-formel:2}
P_m(t,T) := \IE_{m,t}\left[\exp\left(-\int_t^T h Y_s \, ds \right)\right], \quad m \in \{0,1\}.
\ee
Using (\ref{zb-formel:2}), we rewrite (\ref{zb-formel:1}) as follows:
\be \label{zb-formel:3}
 P(t,T) = e^{-r_m(T-t)}\cdot P_1(t,T)^{k} \cdot P_0(t,T)^{N-k}.
\ee 
From the power series representation of the exponential function, we obtain
\bea \label{zb-integral1}
P_m(t,T) &=& \IE_{m,t}\left[1 + \sum_{n=0}^{\infty} \frac{1}{n!} \left(-\int_t^T h Y_s \, ds \right)^n \right] \\
  &=& 1 + \sum_{n=0}^{\infty} (-1)^n h^n 
      \int_t^T\int_{s_1}^T\dots\int_{s_{n-1}}^T \IE_{m,t} [Y_{s_n}\cdots Y_{s_1}]\, ds_n\dots ds_2ds_1, \nonumber
\eea
where $t =: s_0 < s_1 < \dots < s_n < T.$ The last equality follows from
\beo
  \left(\int_t^T Y_s \, ds \right)^n 
      = n!\int_t^T\int_{s_1}^T\dots\int_{s_{n-1}}^T Y_{s_n}\cdots Y_{s_1}\, ds_n\dots ds_2ds_1
\eeo
and the dominated convergence theorem.

For the given $t < s_1 < \dots < s_n < T$ and $t_j := s_j-s_{j-1}$ we have
\beo
 \IE_{m,t} [Y_{s_n}\cdots Y_{s_1}] = 
 \sum_{m_1=0}^1 \cdots \sum_{m_n=0}^1 \prod_{j = 1}^n m_j p_{m_{j-1},m_j}(t_j),
\eeo
where $m_0 := m.$ Using the symmetry relation (\ref{k_symmetry}), we write the transition probabilities $p_{m_{j-1},m_j}(t_j)$ given in Theorem \ref{trm-e-process} as follows:
\beo
  p_{m_{j-1},m_j}(t_j) = w(m_j) \sum_{i_j = 0}^{1} \pi_{i_j} K_{m_{j-1}}(i_j) K_{m_j}(i_j)
   e^{-\lambda_{i_j} t_j },
\eeo
where $\lambda_{i_j} := \lambda(\alpha+\beta) i_j.$ Analogue to the calculation of the expected value (\ref{e-calc1}) in the proof of 
Theorem \ref{trm-e-process}, we obtain
\beo
 \IE_{m_{n-1},s_{n-1}} [Y_{s_n}] 
%   &=& \sum_{y_n=0}^1 y_n p_{y_{n-1},y_n}(t_n) \\
%   &=& \sum_{y_n=0}^1 y_n w(y_n) \sum_{i_n = 0}^{1} \pi_{i_n} K_{y_{n-1}}(i_n) K_{y_n}(i_n)
%       e^{-\lambda_{i_n} t_n } \\
%   &=& \sum_{i_n=0}^1 \pi_{i_n} K_{y_{n-1}}(i_n) e^{-\lambda_{i_n} t_n } \cdot
%       \sum_{y_n=0}^1 y_n K_{i_n}(y_n) w(y_n) \\
   = \sum_{i_n=0}^1 \pi_{i_n} B_{i_{n-1},i_n} K_{m_{n-1}}(i_n) e^{-\lambda_{i_n} t_n},    
\eeo
where $B_{i_{n-1},i_n}$ is defined by (\ref{k_ortho-recurrence}) and $i_0 := 0.$ Iteratively, we get
\beo
 \IE_{m,t} [Y_{s_n}\cdots Y_{s_1}] =
   \sum_{i_1=0}^1 \cdots \sum_{i_n=0}^1 K_m(i_n) \left[\prod_{j =1}^n \pi_{i_j} B_{i_{j-1},i_j} 
   e^{-\lambda_{i_j} t_{j}}\right]. 
\eeo  
Hence, (\ref{zb-integral1}) becomes
\bea \label{zb-integral2}
P_m(t,T) &=& 1 + \sum_{n=0}^{\infty} (-1)^n h^n 
               \sum_{i_1=0}^1 \cdots\sum_{i_n=0}^1 K_m(i_n) \left[\prod_{j =1}^n \pi_{i_j} B_{i_{j-1},i_j}\right] \\
     && \qquad \int_t^T\int_{s_1}^T\dots\int_{s_{n-1}}^T
        \exp\left(-\sum_{k = 1}^n \lambda_{i_k} (s_{k}- s_{k-1} )\right)\, ds_n\dots ds_2ds_1.  \nonumber
\eea
In order to evaluate the multiple integrals above, we tranform the integration domain to the standard simplex
$\Delta_n$ defined by (\ref{def-simplex}) via the following mapping:
\be \label{trafo}
 J : \IR^n \longrightarrow \IR^n,\ 
 \begin{pmatrix} s_1 \\ s_2 \\ \vdots \\ s_n \end{pmatrix}  \longmapsto 
 \begin{pmatrix} (T-t)s_1 + t \\ (T-t)(s_1+s_2) + t \\ \vdots \\ (T-t)(s_1+\cdots +s_n) + t \end{pmatrix}.
\ee
Using (\ref{zb-integral2}), we rewrite (\ref{trafo}) as follows: 
\bea \label{zb-integral3}
P_m(t,T) &=& 1 + \sum_{n=0}^{\infty} (-1)^n h^n 
               \sum_{i_1=0}^1 \cdots\sum_{i_n=0}^1 K_m(i_n) \left[\prod_{j =1}^n \pi_{i_j} B_{i_{j-1},i_j}\right] \\
     && \ \quad\qquad \cdot(T-t)^n \int_{\Delta_{n}} e^{\langle z^{(n)},x \rangle\ }\, dx,  \nonumber
\eea
where $z^{(n)} := -(T-t)(\lambda_{i_1},\cdots,\lambda_{i_n})^T \in \IR^n$ and $\left\langle \cdot\, , \cdot \right\rangle$ denotes the standard inner product. Applying (\ref{simplex-trafo-allg}) with $a = 1,$ we express the integrals in (\ref{zb-integral3}) as $_1F_1$ functions as follows:
\be\label{hyp:1}
 n! \int_{\Delta_n} e^{\langle z^{(n)},x\rangle}\ dx =\, _1F_1\left(1;n+1;z^{(n)}\right)
\ee 
for all $z^{(n)} \in \IR^n \ (n \in \IN_0),$ setting $_1F_1(1;1;z^{(0)}) := 1.$ 
If we combine (\ref{zb-formel:3}) and (\ref{zb-integral3}) with (\ref{hyp:1}) and the fact that $\pi_{i_j} = \left(\frac{p}{q}\right)^{i_j}, \ i_j \in \{0,1\},$ the theorem follows.
\qed

Now we consider the $\Eab$ model with $\alpha=\beta =1$. 
%i.e. the underlying process is the \textit{one-parameter Ehrenfest process}. 
On the one hand, we lose one of the fitting parameters, although, the model is still well suited to model the term structure, and it yields the famous \textit{Vasicek model} in the limit (see Section \ref{sec:vasicek}). On the other hand, we obtain a more tractable pricing formula for ZCBs, where, in contrast to the general case, no multiple sums need calculation, which improves the computational speed.% and hence the accuracy.

The calculation of the arbitrage-free ZCB price (\ref{zb-price}) in this setting is very intuitive and requires no knowledge of the transition probabilities of $(R_t)_{t \geq 0},$ since the only stochastic parameters are the arrival times of the underlying Poisson process.% (see (\ref{k-matrix}) and (\ref{substitution-formula})).
%%%%%%%%%%%%%%%%%%%%%%%%%%%%%%%%%%%%%%%%%%%%%%%%%%%%%%%%%%%%%%%%%%%%%%%%%%%%%%%%%%%%%%%  
\begin{theorem}[ZCB price in $\Eab$ model] \label{trm:zbformel:1}
Let $(R_t)_{t \geq 0}$ be given by the definition (\ref{zinsprocess:1}) with $\lambda > 0$ and $\alpha = \beta = 1.$ The price at time $t\geq 0$ of a ZCB with maturity at $T$ is given by
\be \label{zb:ehrenfest}
P(t,T) = e^{-(r_m +\, \lambda N)(T-t)}\cdot P_1(t,T)^k \cdot P_0(t,T)^{N-k},
\ee
where $k := \frac{R_t - r_m}{h} \in \{0,\dots,N\},$
\bea \label{zb_1}
P_1(t,T) &:=& \sum_{n=0}^{\infty} \frac{(\lambda (T-t))^{2n}}{(2n)!}\cdot       
             \Big\{e^{-h(T-t)}\cdot\, _1F_1\left(1;2n+1;z^{(2n)}\right) \\
             && \qquad \qquad \quad \ \ \, + \ \frac{\lambda (T-t)}{2n+1}\cdot\,
              _1F_1\left(1;2n+2;-z^{(2n+1)}\right) \Big\}, \nonumber \\
\label{zb_0}              
P_0(t,T) &:=& \sum_{n=0}^{\infty} \frac{(\lambda (T-t))^{2n}}{(2n)!}\cdot\, 
             \Big\{\, _1F_1\left(1;2n+1;-z^{(2n)}\right)  \\
             &&  \qquad  \ \ \, + \ 
             \frac{\lambda (T-t)}{2n+1}e^{-h(T-t)}\cdot\, _1F_1\left(1;2n+2;z^{(2n+1)}\right) \Big\},\nonumber
\eea 
where $z^{(2n)} := h(T-t)(0,1,\dots,0,1)^T \in \IR^{2n},\ z^{(2n+1)} := h(T-t)(1,0,1,\dots,0,1)^T \in \IR^{2n+1},$
and $\, _1F_1$ is defined by (\ref{hyp-formel-gen}).
\end{theorem}
\proof
%%%%%%%%%%%%%%%%%%%%%%%%%%%%%%%%%%%%%%%%%% Step 1%%%%%%%%%%%%%%%%%%%%%%%%%%%%%%%%%%%%%%%%%%%%%%%%%%%%%%%%%%%%%%%%%%%%
Let $r_k = hk + r_m, \ k \in \{0,\dots,N\},$ be the state of $(R_t)_{t\geq 0}$ at time $t.$
Analogue to the derivation of the expresion (\ref{zb-formel:3}) in the proof of Theorem \ref{trm:zbformel:p}, we obtain
\be \label{zb-f3}
 P(t,T) = e^{-r_m(T-t)}\cdot \tilde{P}_1(t,T)^{k} \cdot \tilde{P}_0(t,T)^{N-k},
\ee
where
\be \label{zb-f4}
\tilde{P}_y(t,T) := \IE_{y,t}\left[\exp\left(-\int_t^T h Y_s \, ds \right)\right], \quad y \in \{0,1\}.                       
\ee
% Substituting 
%$(R_t)_{t\geq 0}$ due to (\ref{substitution-formula}) and (\ref{zinsprocess:1}) we get from (\ref{zb-price})
%\bea \label{zb-formel:1}
%P(t,T) %&=& \IE\left[\exp\left(-\int_t^T R_s\, ds \right) \Bigg| R_t = r\right] \nonumber \\
% &=& \IE\left[\exp\left(-\int_t^T \left( h\sum_{i=1}^{N} Y^i_s + r_m \right)\, ds \right) \Bigg| X_t = k\right]\nonumber\\
% &=& e^{-r_m(T-t)}\cdot\IE\left[\prod_{i=1}^{N} \exp\left(-\int_t^T h Y^i_s \, ds \right) \Bigg| X_t = k\right] \nonumber\\
% &=& e^{-r_m(T-t)}\cdot\IE\left[\exp\left(-\int_t^T h Y^1_s \, ds \right) \Bigg| Y^1_t = 1\right]^k
%               \cdot\IE\left[\exp\left(-\int_t^T h Y^1_s \, ds \right) \Bigg| Y^1_t = 0\right]^{N-k} 
    %&=& e^{-r_m(T-t)}\cdot \tilde{P}^{(N)}_1(t,T)^{k} \cdot \tilde{P}^{(N)}_0(t,T)^{N-k},             
%\eea
%The last equality holds because of the indipendence and identical distribution of $(Y^i_t = \hat{Y}_{N^i_t})_{t\geq 0}$ for %all 
%$i = 1,\dots,N.$ In the following we omit writing the dependence on particular $i$ and set
%\be \label{zb-formel:2}
%\tilde{P}_j(t,T) := \IE\left[\exp\left(-\int_t^T h Y_s \, ds \right) \Big| Y_t = j\right], \quad j \in \{0,1\}.
%\ee
%Expressing (\ref{zb-formel:1}) due to (\ref{zb-formel:2}), we obtain
%\be \label{zb-formel:3}
% P(t,T) = e^{-r_m(T-t)}\cdot \tilde{P}_1(t,T)^{k} \cdot \tilde{P}_0(t,T)^{N-k}.
%\ee
%%%%%%%%%%%%%%%%%%%%%%%%%%%%%%%%%%%%%%%%%% Step 2%%%%%%%%%%%%%%%%%%%%%%%%%%%%%%%%%%%%%%%%%%%%%%%%%%%%%%%%%%%%%%%%%%%%
In order to evaluate $\tilde{P}_y(t,T)$, we count the number of jumps in the underlying Poisson process $(N_t)_{t\geq 0}$
within the time interval $(t,T],$ and, denoting the jump times by $(\tau_i)_{i\in \IN}$ and setting $\tau_0 := t,$
we split the integral on the right-hand side of (\ref{zb-f4}), obtaining
\beo %\label{zb-formel:4}
 \tilde{P}_y(t,T) 
 %&=& \IE\left[\exp\left(-\int_t^T h \hat{Y}_{N_s} \, ds \right) \Bigg| \hat{Y}_{N_t} = 1\right] \\
 = \IE_{y,t}\left[\sum_{n=0}^{\infty} \IT \cdot
     \exp\left(-\sum_{i=0}^{n-1} \int_{\tau_i}^{\tau_{i+1}} h \hat{Y}_i \, ds - 
     \int_{\tau_n}^T h \hat{Y}_n \, ds \right)\right].
\eeo
At this point, we have to distinguish between even and odd numbers of jumps, since $(\hat{Y}_n)_{n\in \IN}$
switches between 0 and 1 $\IP-$a.s. according to its transition probability matrix (\ref{k-matrix}).
Thus, conditional on $\{\hat{Y}_0 = 1\},$ the Markov chain $(\hat{Y}_n)_{n \in \IN}$ stays in 1 after an even jump, whereas it stays in 0 after an odd jump. This consideration yields
\beao     
\tilde{P}_1(t,T)  &=& \sum_{n=0}^{\infty} \Bigg\{ \IE\left[ \ITT \cdot \exp\left(-\sum_{i=0}^{n-1} \int_{\tau_{2i}}^{\tau_{2i+1}} h \, ds - 
     \int_{\tau_{2n}}^T h \, ds \right) \right] \\
 &&\quad \ \,  +\ \IE\left[ \ITTT \cdot \exp\left(-\sum_{i=0}^n \int_{\tau_{2i}}^{\tau_{2i+1}} h \, ds \right) \right]\Bigg\}\\
% &=& \sum_{n=0}^{\infty} \Big\{ \IE\left[ \exp\left( h(t - \tau_1 + \tau_2 - \tau_3 + \dots
%      + \tau_{2n} - T) \right) \Big| N_{\tau} = 2n \right]\cdot \mathbb{P} \left( N_{\tau}=2n \right) \\
% &&\quad \ \,  +\ \IE\left[ \exp\left( h(t - \tau_1 + \tau_2 - \tau_3 + \dots
%      + \tau_{2n} - \tau_{2n+1}) \right) \Big| N_{\tau} = 2n+1 \right]\cdot \mathbb{P}\left( N_{\tau}=2n+1 \right) \Big\} \\  
\eeao
\beao
\phantom{\tilde{P}_1(t,T)} 
    &=& \sum_{n=0}^{\infty} \Bigg\{ \IE\left[ \exp\left( h\cdot \sum_{i=1}^{2n} (-1)^i\tau_i + t - T \right) \Bigg| 
   N_{T-t} = 2n \right]\cdot \mathbb{P} \left( N_{T-t}=2n \right) \\
 &&\quad \  +\ \IE\left[ \exp\left( h\cdot \sum_{i=1}^{2n+1} (-1)^i\tau_i + t \right) \Bigg| N_{T-t} = 2n+1 \right]\cdot \mathbb{P}\left( N_{T-t}=2n+1 \right) \Bigg\}. 
\eeao
Furthermore, from the order statistics property of the Poisson process (see, for example, \cite{Ref_Karlin_2}, pp. 101-102),
we know that the joint density of the arrival times 
$\tau_1,\dots,\tau_k \ (k\in \IN)$ of $(N_t)_{t \geq 0}$ in $(t,T],$ conditional on $\{N_{T-t} = k\},$ is given by
\be
 \mathbb{P} (t < \tau_1 \leq \tau_2 \leq \dots \leq \tau_k \leq T | N_{T-t} = k) = 
  \frac{n!}{(T-t)^k} \int_t^T \int_{t_1}^T \dots \int_{t_{k-1}}^T \ dt_k\dots dt_2dt_1.
\ee 
Hence, for $\tilde{P}_1(t,T)$ we obtain 
\bea \label{zb-formel:4}
 \tilde{P}_1(t,T) &=& \sum_{n=0}^{\infty}\Bigg\{ 
   \left[ e^{-h(T-t)} \cdot \frac{(2n)!}{(T-t)^{2n}}  
   \int_t^T \dots \int_{t_{2n-1}}^T \exp\left( h\cdot \sum_{i=1}^{2n} (-1)^i t_i \right) \ dt_{2n}\dots dt_1 \right] \nonumber\\
&& \qquad \ \ \cdot\ e^{-\lambda (T-t)} \frac{(\lambda (T-t))^{2n}}{(2n)!} \nonumber\\
&&\quad \ \,  + \ \left[ e^{-ht} \cdot \frac{(2n+1)!}{(T-t)^{2n+1}} 
   \int_t^T \dots \int_{t_{2n}}^T \exp\left( h\cdot \sum_{i=1}^{2n+1} (-1)^i t_i \right) \ dt_{2n+1}\dots dt_1 \right] \nonumber\\
&& \qquad \ \ \cdot\ e^{-\lambda (T-t)} \frac{(\lambda (T-t))^{2n+1}}{(2n+1)!}  \Bigg\}. 
\eea
%In order to evaluate the integrals above, we tranform the integration domain to the standard simplex
%in $\IR^d\ (d\in \IN)$ defined by
%\beo
% \Delta_d := \left\{(x_1,\dots,x_d)\in \IR^d :x_i \geq 0, i=1,\dots d, \sum_{i=1}^d x_i \leq 1 \right\}
%\eeo 
%via the following mapping
%\be \label{trafo}
% J : \IR^d \longrightarrow \IR^d,\ 
% \begin{pmatrix} t_1 \\ t_2 \\ \vdots \\ t_d \end{pmatrix}  \longmapsto 
% \begin{pmatrix} (T-t)t_1 + t \\ (T-t)(t_1+t_2) + t \\ \vdots \\ (T-t)(t_1+\cdots +t_d) + t \end{pmatrix}.
%\ee
%For the sake of convinience, we define $z^{(2n)} := (0,1,\dots,0,1) \in \IR^{2n} \text{ and } z^{(2n+1)} := (1,0,1,\dots,0,1) %\in \IR^{2n+1}.$ 
Analogue to the proof of Theorem \ref{trm:zbformel:p}, we consider the mapping $J$ given in (\ref{trafo})
and integrate (\ref{zb-formel:4}) by substitution, which yields 
\bea \label{zb-formel:5}
\tilde{P}^{(N)}_1(t,T) &=& e^{-\lambda(T-t)} \sum_{n=0}^{\infty}\frac{(\lambda (T-t))^{2n}}{(2n)!}
 \Bigg\{ e^{-h(T-t)} \cdot (2n)! \int_{\Delta_{2n}} e^{ \langle z^{(2n)},x \rangle\ } dx \nonumber\\
  &&  \qquad \quad \qquad \qquad \qquad \ \, + \ (2n+1)! \int_{\Delta_{2n+1}} e^{\langle -z^{(2n+1)},x\rangle\ }  
      dx \Bigg\},
\eea
where $z^{(2n)} := h(T-t)(0,1,\dots,0,1)^T \in \IR^{2n},\ z^{(2n+1)} := h(T-t)(1,0,1,\dots,0,1)^T \in \IR^{2n+1},$ and 
$\left\langle \cdot\, , \cdot \right\rangle$ denotes the standard inner product.
Applying the relation (\ref{hyp:1}), we rewrite (\ref{zb-formel:5}) as follows:
\bea \label{zb-formel:6}
 \tilde{P}_1(t,T) &=& e^{-\lambda(T-t)}\sum_{n=0}^{\infty}\frac{(\lambda (T-t))^{2n}}{(2n)!}
  \Big\{ e^{-h(T-t)}\cdot \, _1F_1\left(1;2n+1;z^{(2n)}\right) \nonumber\\
  &&\qquad \qquad \qquad \qquad \quad \ \, + \ \frac{\lambda (T-t)}{2n+1}\cdot\,
   _1F_1\left(1;2n+2;-z^{(2n+1)}\right) \Big\} \nonumber\\
  &=:& e^{-\lambda(T-t)} P_1(t,T). 
\eea
In similar fashion, we obtain
\bea \label{zb-formel:7}
 \tilde{P}_0(t,T) &=& e^{-\lambda(T-t)}\sum_{n=0}^{\infty}\frac{(\lambda (T-t))^{2n}}{(2n)!}
  \Big\{\, _1F_1\left(1;2n+1;-z^{(2n)}\right) \nonumber\\
  &&  \qquad \qquad \quad \ \ \, + \ \frac{\lambda (T-t)}{2n+1} e^{-h(T-t)}\,
   _1F_1\left(1;2n+2;z^{(2n+1)}\right) \Big\} \nonumber\\
  &=:& e^{-\lambda(T-t)} P_0(t,T). 
\eea
If we combine (\ref{zb-formel:6}) and (\ref{zb-formel:7}) with (\ref{zb-f3}), the theorem follows.
\qed
\begin{remark}
We notice that Theorems \ref{trm:zbformel:p} and \ref{trm:zbformel:1} are based on two completely different approaches 
and yield different representations of the ZCB prices. However, both formulae involve the confluent hypergeometric function $_1F_1$ defined by (\ref{hyp-formel-gen}). 
\end{remark}
%It is remarkable that zero-coupon bond prices in the general and the special cases, albeit proved differently, 
%are both expressed in terms of weighted sums of hypergeometric functions of matrix argument.  
%%%%%%%%%%%%%%%%%%%%%%%%%%%%%%%%%%%%%%%%%%%%%%%%%%%%%%%%%%%%%%%%%%%%%%%%%%%%%%%%%%%%%%%%%%%%%%%%%%%%%%%%%%%%%%%%%
\paragraph{Practical implementation}
From Theorems \ref{trm:zbformel:p} and \ref{trm:zbformel:1}, the ZCB prices can be computed approximately by
truncating the series in the according formulae. We also use the truncated $_1F_1^H$ function defined
by (\ref{hyp-trunvated}) as an approximation for the $_1F_1$ function. 

Thus, in the setting of Theorem \ref{trm:zbformel:p},
%in the general case of the Ehrenfest short-rate model with $\alpha, \beta \in (0,1],$ 
we truncate the sum of series (\ref{zb-series1}), obtaining 
\bea\label{zb:trunvated_gen}
P(t,T;M,H) &:=& e^{-r_m(T-t)}\cdot P_1(t,T;M,H)^k \cdot P_0(t,T;M,H)^{N-k}, \\
P_y(t,T;M,H) &:=& 1 + \sum_{n=1}^{M} (-1)^n \frac{(-h(T-t))^n}{n!} \cdot \nonumber\\
         &&\sum_{i_1 = 0}^1 \cdots \sum_{i_n = 0}^1 K_y(i_n)
            \left[\prod_{j =1}^n \left(\frac{p}{q}\right)^{i_j}B_{i_{j-1},i_j}\right]  \cdot
            \, _1F_1^H\left(1;n+1;z^{(n)}\right) \nonumber          
\eea 
for $y \in \{0,1\}.$ 
%for $y \in \{0,1\},$ where $z^{(n)} := -\lambda(\alpha+\beta)(T-t)(i_1,\dots,i_n)^T \in \IR^{n},\ i_0 := 0,\ 
%K_y$ are given due to $(b)$ and $(c)$ of Lemma \ref{k-lemma}, $\, _1F_1$ is defined by (\ref{hyp-formel-gen}) and
%\beo
% B_{i_{j-1},i_j} := \begin{cases}
%                      i_j(p-1) \quad & \text{if}\quad i_{j-1} = i_j-1,\\
%                      i_j(1-2p) + p \quad & \text{if}\quad i_{j-1} = i_j.
%                   \end{cases}
%\eeo

In the setting of Theorem \ref{trm:zbformel:1}, we truncate the series (\ref{zb_1}) and (\ref{zb_0}), obtaining 
%an approximation for the ZCB the special case of the Ehrenfest short-rate model with $\alpha=\beta=1$ as fo
\bea \label{zb:trunvated_spec}
P(t,T;M,H) &:=& e^{-(r_m + a)(T-t)}\cdot P_1(t,T,M,H)^k \cdot P_0(t,T,M,H)^{N-k}, \\
P_1(t,T;M,H) &:=& \sum_{n=0}^{M} \frac{(\lambda (T-t))^{2n}}{(2n)!}\cdot       
             \Big\{e^{-h(T-t)}\cdot\, _1F_1^H\left(1;2n+1;h(T-t) z^{(2n)}\right) \nonumber\\
             && \qquad \qquad \quad \ \ \, + \ \frac{\lambda (T-t)}{2n+1}\cdot\,
              _1F_1^H\left(1;2n+2;-h(T-t) z^{(2n+1)}\right) \Big\},  \nonumber\\             
P_0(t,T;M,H) &:=& \sum_{n=0}^{M} \frac{(\lambda (T-t))^{2n}}{(2n)!}\cdot\, 
             \Big\{\, _1F_1^H\left(1;2n+1;-h(T-t) z^{(2n)}\right)  \nonumber\\
             &&  \qquad  \ \ \, + \ 
             \frac{\lambda (T-t)}{2n+1}e^{-h(T-t)}\cdot\, _1F_1^H\left(1;2n+2;h(T-t) z^{(2n+1)}\right) \Big\}. \nonumber
\eea 
%and $\lambda, z^{(2n)} \text{ and } z^{(2n+1)}$ as in Theorem \ref{trm:zbformel:1}. 
The choice of the truncation parameters $M$ and $H$ is left to the practitioner and should be made in the way of
maintaining a balance between the accuracy of the results and the computational speed. Some numerical examples, which provide numerical accuracy and computational speed for the formulae (\ref{zb:trunvated_gen}) and (\ref{zb:trunvated_spec}), will be given at the end of the next section. In the following we omit explicitly writing out the dependence on $M$ and $H$ 
when it is clear from the context. 
%%%%%%%%%%%%%%%%%%%%%%%%%%%%%%%%%%%%%%%%%%%%%%%%%%%%%%%%%%%%%%%%%%%%%%%%%%%%%%%%%%%%%%%%%%%%%%%%%%%%%%%%%%%%%%%%%%%%%%%%
\section{Connection to the Vasicek model}
\label{sec:vasicek}
The Vasicek model \cite{Ref_Vasicek} is one of the most popular short-rate models. Closed-form expressions of ZCB prices
and European options thereof make the model highly appealing to practitioners. However, it also has some shortcomings.
 
In this section we give a short description of the Vasicek model. We point out its advantages and disadvantages. Here we follow
paragraph 3.2.1 of \cite{Ref_Brigo}. We provide a convergence result, which shows that after a linear rescaling the $\Eab$ model converges weakly to the Vasicek model. We show the convergence of the respective ZCB prices and provide
some numerical examples.% examining the accuracy of the pricing formulae
\paragraph{Vasicek model}
The formulation of the Vasicek model under the risk-neutral measure $\IP$ is
\be \label{sdgl:vasicek}
 dr(t) = k[\theta - r(t)]dt + \sigma dW_t, \quad r(0) = r_0,
\ee
where $k, \theta, \sigma, r_0$ are positive constants, and $(W_t)_{t \geq 0}$ is the
standard Wiener process on the probability space $(\Omega, \IF,(\IF_t)_{t\geq 0 }, \IP)$ with the
natural filtration $(\IF_t)_{t\geq 0 }.$ 
Integration of the equation (\ref{sdgl:vasicek}) yields for $t \geq s$
\be\label{vasicek-process}
r(t) = r(s)e^{-k(t-s)} + \theta \left(1-e^{-k(t-s)} \right) + \sigma\int_s^t e^{-k(t-u)}\, dW_u.
\ee
Thus, conditional on $\IF_s\ (t \geq s),\ r(t)$ is normally distributed with mean and variance
\bea\label{vasicek-distr}
 \IE [r(t)| \IF_s] &=& r(s)e^{-k(t-s)} + \theta\left(1-e^{-k(t-s)}\right), \\
 \IV ar[r(t)| \IF_s] &=& \frac{\sigma^2}{2k}\left(1-e^{-2k(t-s)}\right).
% r(t) | \IF_s \sim \mathcal{N} \left( r(s)e^{-k(t-s)} + \theta(1-e^{-k(t-s)}),\frac{\sigma^2}{2k}(1-e^{-2k(t-s)}) \right).
\eea
Hence, the short-rate process $(r_t)_{t\geq 0}$ tends to the mean-reverting value $\theta$ for $t \rightarrow \infty.$
The drawbacks of the model are the possible negativity of the interest rates, implied by the Gaussian distribution,
and the fact that it is driven by only three parameters, which makes the calibration an ill-posed problem and yields often poor results.
   
The price at time $t\geq 0$ of a ZCB with maturity at $T,$ conditional on $r(t) = r,$ is given by
\be \label{zb:vasicek}
 P(t,T) = A(t,T)e^{-B(t,T)\cdot r},
\ee
where
\beao
 A(t,T) &=& \exp \left\{ \left(\theta - \frac{\sigma^2}{2k^2}\right)[B(t,T) - T + t]
            - \frac{\sigma^2}{4k}B(t,T)^2   \right\}, \\
 B(t,T) &=& \frac{1}{k}\left(1 - e^{-k(T-t)} \right).           
\eeao
%There also exists an explicit formula for European options on ZCBs (see \cite{Ref_Brigo}, p. 60). 
%\subsection{Convergence of the Ehrenfest model to the Vasicek model} \label{e-v-convergence}
\paragraph{Convergence results}
It is well known that the Ehrenfest process converges weakly to the Ornstein-Uhlenbeck process (see, for instance, \cite{Ref_Karlin_2}, pp. 168-173, or \cite{Ref_Sumita}). The following theorem shows that the $\Eab$ model also converges weakly to the Vasicek model.
%%%%%%%%%%%%%%%%%%%%%%%%%%%%%%%%%%%%%%%
\begin{theorem} \label{convegence:1}
Let $(r(t))_{t \geq 0}$ be given as in (\ref{sdgl:vasicek}). Consider $(R^{(N)}_t)_{t\geq 0}$ as defined in (\ref{zinsprocess:1}) with $\alpha = \beta \in (0,1],\ \lambda := \alpha / (\alpha + \beta), \ 
r_m := \theta - \sigma \sqrt{\frac{N}{2a}}$ and $r_M := \theta + \sigma \sqrt{\frac{N}{2a}}.$ Then, 
\beo
 (R^{(N)}_t)_{t \in [0,T]} \Rightarrow (r(t))_{t \in [0,T]} \quad \text{as} \quad N\rightarrow \infty,
\eeo
where $``\Rightarrow"$ denotes the weak convergence.
\end{theorem}
\proof 
The proof follows analogue to \cite{Ref_Karlin_2}, pp. 168-173, when we compute the conditional moments
of $\Delta R_t := R_{t+\Delta t} - R_t.$ 
%The proof follows after the computation of the conditional moments of $\Delta R_t := R_{t+\Delta t} - R_t$
%(see \cite{Ref_Karlin_2}, pp. 168-173 for details).
\qed
\begin{remark}
Karlin and McGregor show in a rigorous way (see \cite{Ref_Karlin_3}, pp. 371-373), that the transition probability function (\ref{e-trans-prob}) of the Ehrenfest process converges locally uniformly to the transition probability function of the Ornstein-Uhlenbeck process as $N \rightarrow \infty,$ which sharpens the above result after a linear transformation of the underlying processes.
\end{remark}
A direct consequence of Theorem \ref{convegence:1} is the convergence of the respective ZCB prices.
\begin{corollary} \label{convegence:2}
Consider $(r(t))_{t \in [0,T]}$ and $(R^{(N)}_t)_{t \in [0,T]}$ as in Theorem \ref{convegence:1} 
and let $P(t,T)$ and $P^{(N)}(t,T)$ denote the associated ZCB prices at time $t$ with maturity at $T$ given by (\ref{zb:vasicek}) and (\ref{zb:ehrenfest}) respectively. Then,
\beo
 P^{(N)}(t,T) \stackrel{N \rightarrow \infty}{\longrightarrow} P(t,T). 
\eeo 
\end{corollary}
\proof
%The proof is clear from Theorem 25.12 in \cite{Ref_Billingsley}.
W.l.o.g. let $t = 0.$ We denote by $\mathbb{D} := \mathbb{D}[0,T]$ the space of the real-valued functions on $[0,T]$
that are \textit{right continuous and have left-hand limits (RCLL)}. From \cite{Ref_Billingsley_1} (see p. 123), we know
that a metric exists that makes $\mathbb{D}$ a \textit{Polish space}, i.e. a metric, separable and complete space. Clearly, $R_N := (R^{(N)}_t)_{t \in [0,T]}$ and $r := (r(t))_{t \in [0,T]}$ both lie in $\mathbb{D}$. With
Theorem \ref{convegence:1}, it follows that $R_N \Rightarrow r$ in $\mathbb{D}.$ 

Consider a linear operator $S$ on $\mathbb{D},$ defined by
\beo
 (\tilde{S}f)(t) := \int_0^t f(s)\, ds \quad \text{for $f \in \mathbb{D}$ and $t\in [0,T]$}.  
\eeo 
Clearly, $\tilde{S}$ is a continuous operator on $\mathbb{D}.$ Then, the operator $S,$ defined by
\beo
 (Sf)(t) := \exp\left(- (\tilde{S} R_N)(t) \right) =
            \exp\left(- \int_0^t f(s)\, ds \right) \quad \text{for $f \in \mathbb{D}$ and $t\in [0,T]$},
\eeo
is a continuous operator on $\mathbb{D}.$ Let $Y_N := (S R^{(N)}(t))_{t \in [0,T]}$ and 
$Y := (S r(t))_{t \in [0,T]}.$ Then, Theorem 5.1 in \cite{Ref_Billingsley_1} yields $Y_N \Rightarrow Y.$ 
Since $Y_N$ is \textit{uniformly integrable}, it follows from Theorem 5.4 in \cite{Ref_Billingsley_1} that
\beo
\IE \left[\, Y_N  \right] \stackrel{N \rightarrow \infty}{\longrightarrow} \IE \left[\, Y  \right],
\eeo
which completes the proof.
\qed

\begin{figure}[h]
\begin{center}
\subfigure[Favourable case]{\includegraphics[width=.45\linewidth, viewport= 135 275 474 571]{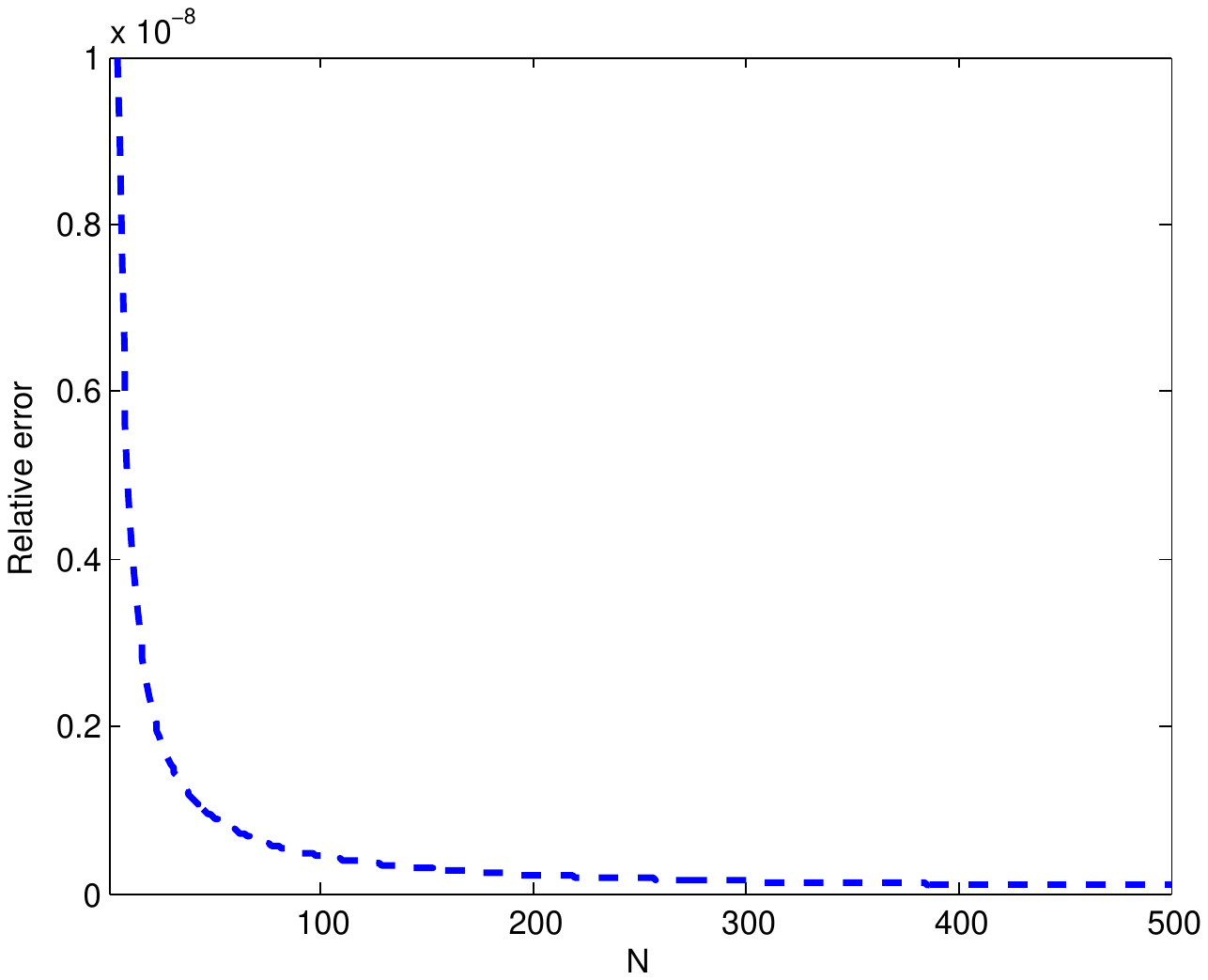}} \quad
\subfigure[Unfavourable case]{\includegraphics[width=.45\linewidth, viewport= 135 275 474 571]{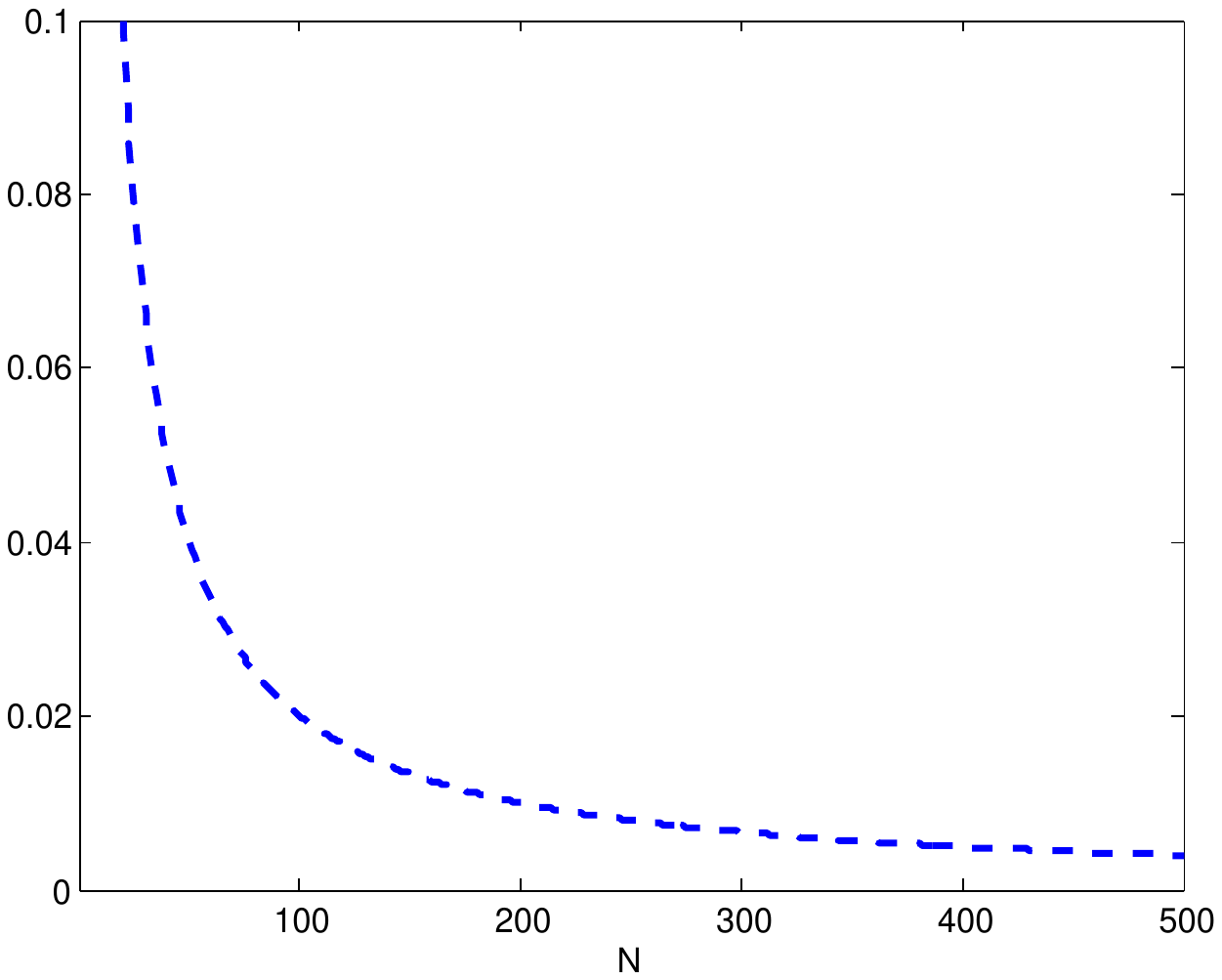}}
\caption{Relative price errors against $N$ approximating ZCB prices $P(0,T)$ in the Vasicek model by $P^{(N)}(0,T;10,30)$ 
in the $\Eab$ model via (\ref{zb:trunvated_spec}) in two scenarios: 
$(a)\ \text{Vasicek model with } a = 0.2,\, \sigma = 0.05,\, T = 1 \text{ year},\, \theta = 8 \%,$ and $r_0 = 5\%$;
$(b)\ \text{Vasicek model with } a = 0.2,\, \sigma = 0.2,\, T = 10 \text{ years},\, \theta = 8\%,$ and $r_0 = 5\%$.}
\label{figure-zb}
\end{center}
\end{figure}

Figure \ref{figure-zb} illustrates the convergence result of Corollary \ref{convegence:2}. All computations were made on an 
INTEL Core2Duo 2400MHz machine. We consider two scenarios: in case $(a)$ we have a favourable set of parameters for the ZCB valuation; in case $(b)$ we choose an unrealistic high of 20\% for the interest rate market volatility and a time to maturity of 10 years. Within the Vasicek model the valuation is done according to the pricing formula (\ref{zb:vasicek}). The approximative values $P(0,T;M,H)$ of the ZCB prices in the $\Eab$ model are computed according to Corallary \ref{convegence:2} via (\ref{zb:trunvated_spec}) with $M = 10$ and $H = 30.$ In both cases we observe fast convergence of the respective prices.

Furthermore, we see that the choice of the truncating parameters $M=10$ and $H=30$ is satisfactory for our purpose.
The computation time of one ZCB price via (\ref{zb:trunvated_spec}) is less than 0.1 seconds.
Similar computation in the $\EAB$ model according to (\ref{zb:trunvated_gen}) takes 2.67 seconds.  
%%%%%%%%%%%%%%%%%%%%%%%%%%%%%%%%%%%%%%%%%%%%%%%%%%%%%%%%%%%%%%%%%%%%%%%%%%%%%%%%%%%%%%%%%%%%%%%%%%%%%%%%%%
\section{Discussion}
\label{sec:discuss}
%Gaussian one-factor short-rate models are highly analytically tractable with respect to the prices of ZCBs and European bond %options, which makes them widely-used in praxis. Though, the most frequently discussed drawback of such models is the %posibility
%of negative interest rates. Albeit with small probability, major problems arise while valuing interest rate derivatives in
%the borderline cases.

In this section we discuss the advantages of the $\EAB$ model with respect to the positivity of the interest rates. 
We use the case study of a ZCB valuation, showing that the $\EAB$ model can still be used when the Vasicek model reaches its limits.     

The main shortcoming of all models with Gaussian distribution, including the Vasicek model, is the positive probability of the interest rates becoming negative. Although this probability is rather small, some problems may appear while valuing
ZCBs with long residual maturity. For instance, Rogers \cite{Ref_Rogers} illustrates how  an attempt to keep the probability of negative interest rates negligible by choosing suitable parameters of the Vasicek model in the limiting case $t \rightarrow \infty$ leads to an exponential growth in $t$ of the ZCB prices. Conversely, the $\EAB$ model allows the choice of
the lower and upper bounds $r_m$ and $r_M$ for the interest rate, and excludes the possibility of negative as well as unrealistically high positive interest rates.    
%The main drawback of the Vasicek model is the positive probability for the interest rate going negativ; however this
%probability is rather small. This shortcoming concerns all models with Gaussian distribution. For instance, Rogers %\cite{Ref_Rogers} illustrates how  an attempt to keep the probability of negative interest rates negligible by choosing %suitible parameters of the Vasicek model in the limiting case $t \rightarrow \infty$ leads to an exponential growth in $t$ of %the ZCB prices.   
%%imposing conditions on the parameters of the Vasicek model in order to prevent negative interest rates leads to absurd.   
%On the contrary, in the Ehrenfest short-rate model we are free to choose the lower and upper bounds $r_m$ and $r_M,$
%excluding the possibility of negative as well as unrealistically high positive interest rates. 

\begin{figure}[h]
\begin{center}
\subfigure[Short-rate]{\includegraphics[width=.45\linewidth, viewport= 135 275 474 571]{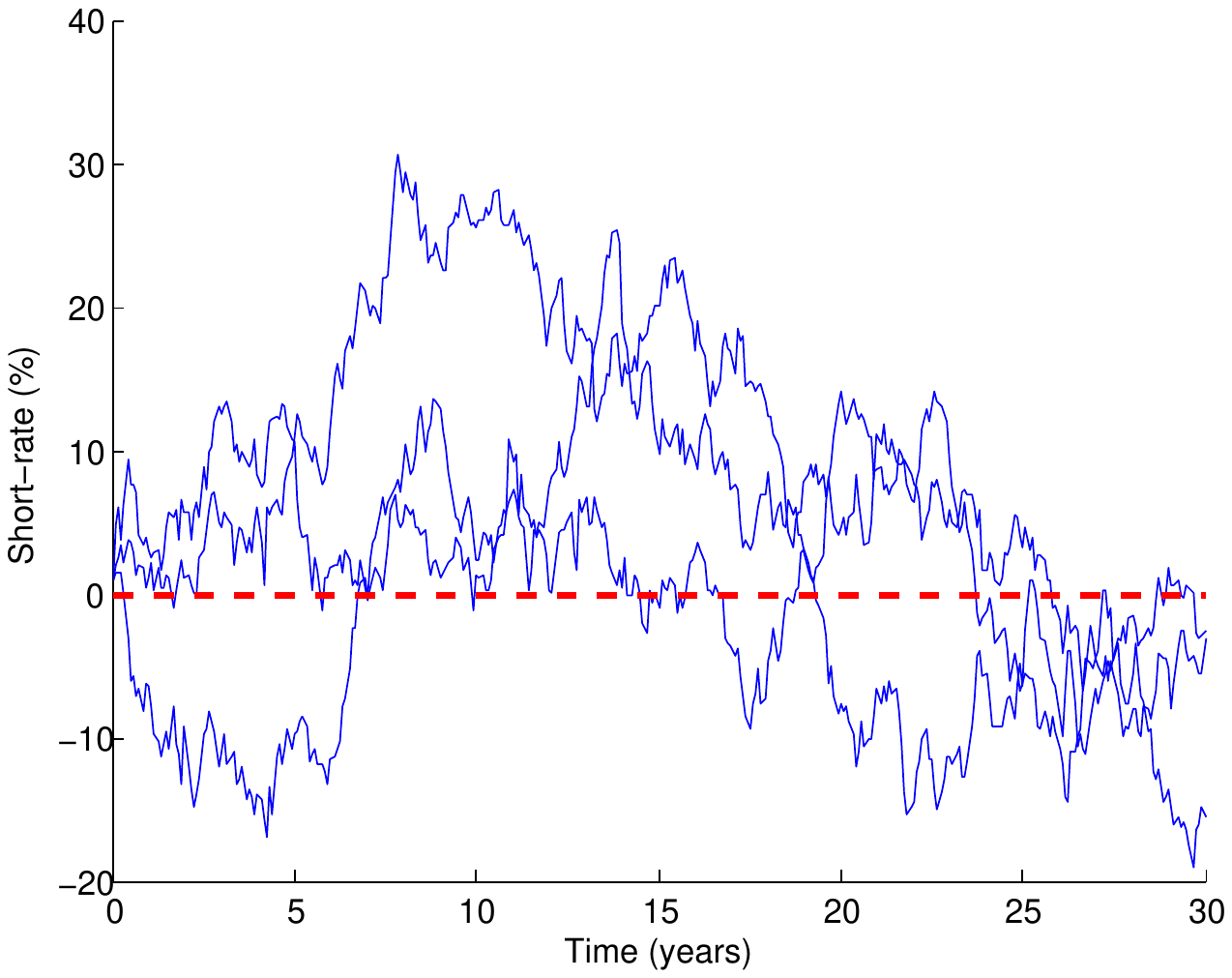}} \quad
\subfigure[ZCB prices]{\includegraphics[width=.45\linewidth, viewport= 135 275 474 571]{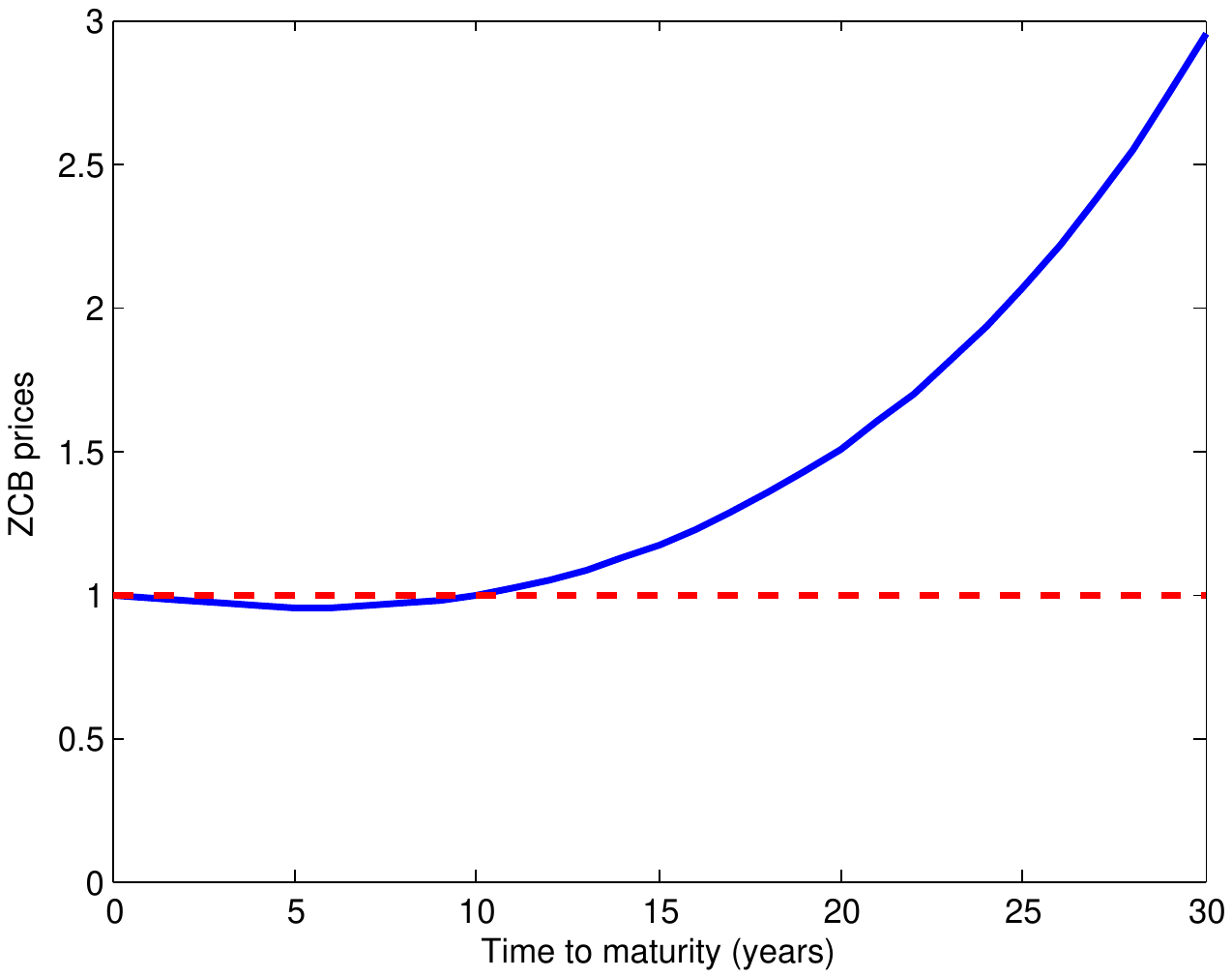}}
\caption{$(a)$ Three sample paths of the short-rate process (\ref{sdgl:vasicek}) in the Vasicek model with $a = 0.1,\, \theta = 4 \%,\, \sigma = 0.05$ and $r(0) = 1 \%.$ $(b)$ ZCB prices in the Vasicek model with the given parameters and residual maturities from 1 to 30 years.}
\label{fig:vasicek}
\end{center}
\end{figure}

Times of financial crisis are often accompanied by interest rates near $0 \%$, as we see at present.
% the mentioned above drwaback can make itself felt. 
The following example of pricing ZCBs in a respective scenario illustrates the advantage of the $\EAB$ model over
the Vasicek model. First, we assume the Vasicek model given according to (\ref{sdgl:vasicek}) with $\theta = 4 \%, \ 
\sigma = 0.05,\ k = 0.1$ and $r(0) = 1 \%.$ Figure \ref{fig:vasicek} $(a)$ shows three sample paths of the underlying process $(r_t)_{t\geq 0}$ over a period of 30 years simulated according to (\ref{vasicek-process}). We see that every path of the simulated process spends some time below the zero mark. Figure \ref{fig:vasicek} $(b)$ demonstrates the weakness of the model in the case at hand, as we observe that the ZCB prices are not monotone falling in the time to maturity and even exceed the upper bound of 1 monetary unit, which is contradictory to no-arbitrage principles.

\begin{figure}[ht]
\begin{center}
\subfigure[Short-rate]{\includegraphics[width=.45\linewidth, viewport= 135 275 474 571]{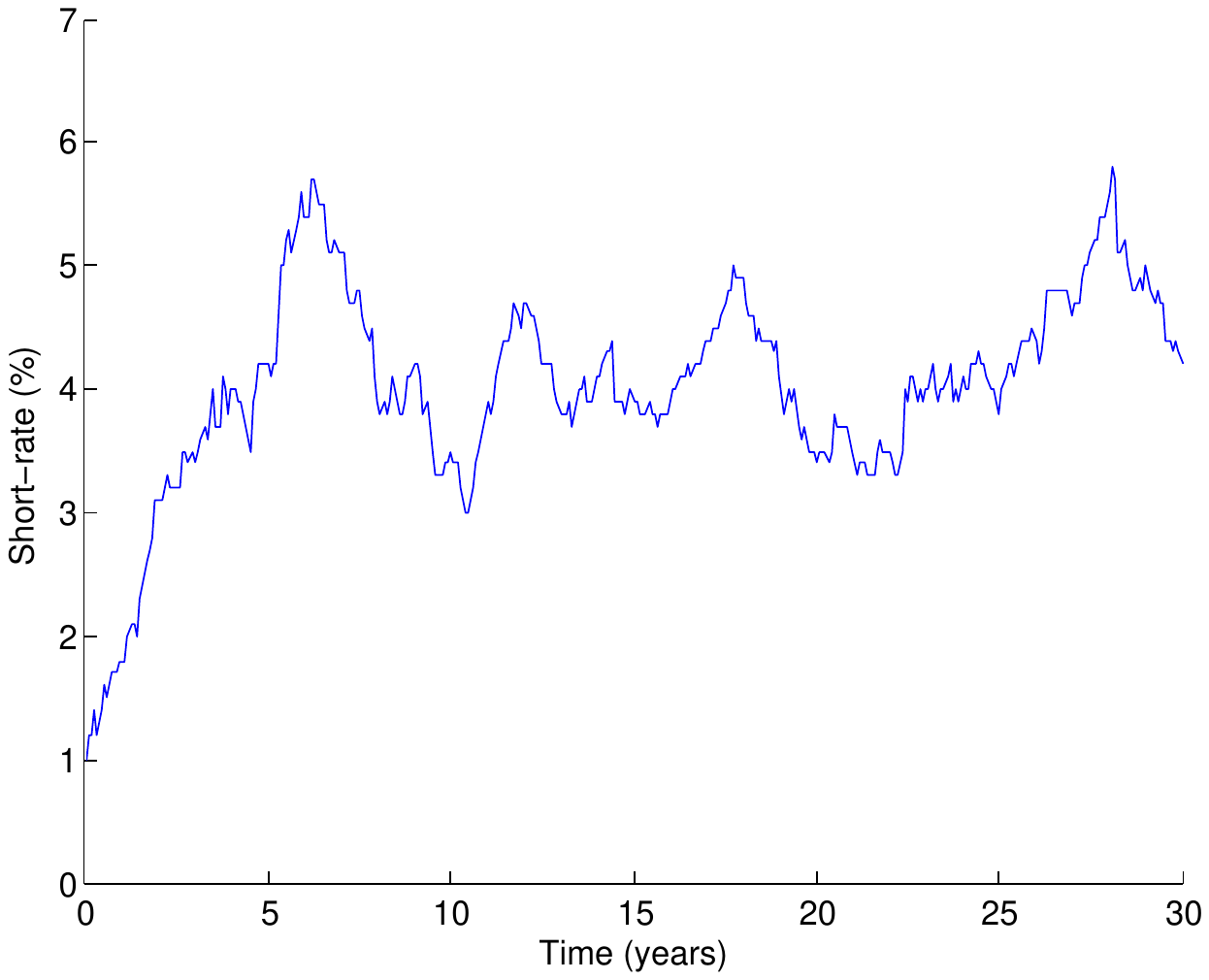}} \quad
\subfigure[ZCB prices]{\includegraphics[width=.45\linewidth, viewport= 135 275 474 571]{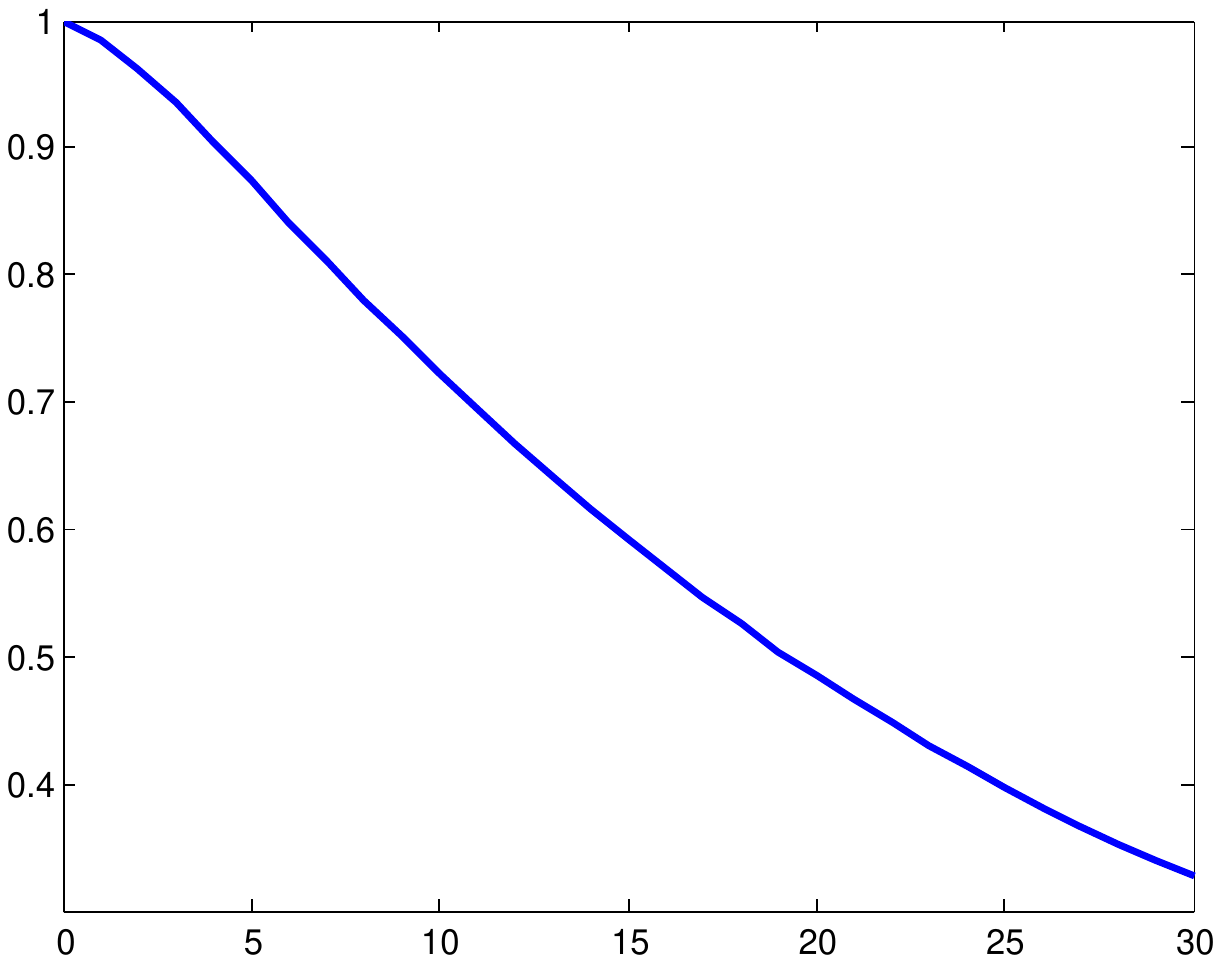}}
\caption{$(a)$ A sample path of the short-rate process (\ref{zinsprocess:1}) in the $\EAB$ model with $\alpha = 0.1,\,
\beta = 0.3,\, \lambda = 1,\, r_m = 0 \%,\, r_M = 16 \%,\, N = 160$ and $R_0 = 1 \%.$ $(b)$ ZCB prices in the $\EAB$ model with the given parameters and residual maturities from 1 to 30 years.}
\label{fig:ehrenfest}
\end{center}
\end{figure}

Now we consider the $\EAB$ model in a similar hypothetical setting. We set the lower and upper bounds at $r_m = 0 \%$ and $r_M = 16 \%$, and the state space discretization parameter $N = 160.$ We choose $\lambda = 1,\, \alpha = 0.1$ and $\beta = 0.3,$ in that we have with (\ref{mean-rev1}) a mean-reverting value of $4 \%$ as in the case above. Here, we set $R_0 = 1 \%$ as well. 
Figure \ref{fig:ehrenfest} $(a)$ demonstrates a possible trajectory of the short-rate process $(R_t)_{t \geq 0}$ over 30 years,
simulated on the basis of the underlying distribution. In Figure \ref{fig:ehrenfest} $(b)$ we see the strictly monotone decreasing character of the respective ZCB prices as a function of the time to maturity, which is highly plausible.  
%%%%%%%%%%%%%%%%%%%%%%%%%%%%%%%%%%%%%%%%%%%%%%%%%%%%%%%%%%%%%%%%%%%%%%%%%%%%%%%%%%%%%%%%%%%%%%%%%%%

\section{Conclusions}
This paper has explored a finite-state mean-reverting short-rate model based on the Ehrenfest process.
The respective short-rate process can be seen as an affine linearly transformed birth-and-death process on 
$\{0,1,\cdots,N\},\ N\in \IN.$ The model provides a certain degree of analytical tractability, since it allows
explicit pricing of ZCBs and solves the problem of negative interest rates characteristic of Gaussian models.
The pricing formulae for ZCBs have been derived for both the general case and the special case, in which the underlying distribution is symmetric with respect to the mean-reverting value. The key to both approaches has turned out to be the representation of the underlying Ehrenfest process as a sum of independent binary processes, which has been possible only in continuous time. We also used the hypergeometric functions of a matrix argument and the Krawtchouk polynomials.
The special case benefits also from a more tractable pricing formula for ZCBs. 

We have seen that the Ehrenfest short-rate model is a good approximation to the Vasicek model under normal conditions
and a better alternative to it in extreme cases, where the interest rates are low and the volatility is high, providing solely positive interest rates. A further advantage of the model is the availability of five fitting parameters in the
general case.

Our conclusion is that especially the general case of the Ehrenfest short-rate model is an interesting enrichment in the field of term structure modelling, combining analytical tractability with the desired property of interest rates remaining positive.  
%
%The main advantage of the proposed model is the free definition of the state space, which, for instance, ensures the %positivity of interest rates. The paper shows that the examined model can be used in 
%
%compares the examined model to the Vasicek model
%Limiting relation to the Vasicek model is shown and respective numerical results are provided.

Problems that remain open for the short-rate model that we have examined here are the derivation of an explicit pricing formula for European options on ZCBs, parameter estimates for the model under the objective measure, and an extension of the model 
according to the three urn Ehrenfest model (see. \cite{Ref_Karlin_3}, pp. 363 - 368).    

\end{document}